\documentstyle[12pt,aaspp]{article}

\def\feh{[Fe/H]\,}
\def\Mrr{\,M$_V$\,}
\def\RR{\,RR~Lyrae\,}
\def\Msol{\,M_\odot}

\def\kms{km\,s$^{-1}$}
\def\vt{$v_{\rm t}$}
\def\Rnc{$R_{NC}$}
\def\Rc{$R_{C}$}
\def\Inc{$I_{NC}$}
\def\Ic{$I_{C}$}
\def\Rbm{$R_{BM}$}
\def\Ibm{$I_{BM}$}
\def\Kcit{$K_{CIT}$}
\def\Kj{$K_{J}$}
\def\teff{$T_{\rm eff}$~}
\def\C12C13{$^{12}$C/$^{13}$C}
\def\DS{$\Delta$S\,}
\def\CaII{Ca II K line\,}

\def\sun{\odot}

\begin{document}
\title{THE COMPOSITION OF HB STARS: RR~LYRAE VARIABLES\altaffilmark{1}}
\altaffiltext{1}{Based in part on data
obtained with the Palomar 1.5m telescope, which is owned and operated
jointly by the California Institute of Technology and the Carnegie
Institute of Washington, and in part on data obtained with the Asiago
Astrophysical Observatory 1.8m telescope}
\altaffiltext{2}{Guest observer, Palomar Mountain Observatory}

\author{G.CLEMENTINI\altaffilmark{2}}
\affil{Osservatorio Astronomico di Bologna, CP 596, I-40126 Bologna}

\author{E. CARRETTA}
\affil{Dipartimento di Astronomia, Universit\'a di Padova, Vicolo
dell'Osservatorio 5, I-35122 Padova and Osservatorio Astronomico di Bologna,
CP 596, I-40126 Bologna}

\author{R. GRATTON}
\affil{Osservatorio Astronomico di Padova, Vicolo dell'Osservatorio 5, I-35122
Padova}

\author{R. MERIGHI}
\affil{Osservatorio Astronomico di Bologna, CP 596, I-40126 Bologna}

\author{J. R. MOULD}
\affil{M.Stromlo and Siding Spring Observatories, Institute of Advanced
Studies, Australian National University, Weston PO, ACT 2611, Australia}

\author{J. K. McCARTHY}
\affil{Palomar Observatory, California Institute of Technology, Pasadena, CA
91125}

\newpage

\begin{abstract}

We have used moderately high-resolution, high S/N spectra to study the chemical
composition of 10 field {\it ab}-type RR~Lyrae stars. Variables having accurate
photometric and radial velocity data were selected, in order to derive a
precise estimate of the atmospheric parameters independently of excitation and
ionization equilibria. A new temperature scale was determined from literature
Infrared Flux Method measures of subdwarfs and the Kurucz (1992) model
atmospheres, and used to calibrate colors for both dwarfs and RR~Lyraes.
Photometric reddening estimates for the program stars were carefully
examined, and compared with other determinations.
The applicability of Kurucz (1992)
model atmospheres in the analysis of RR~Lyraes at minimum light was analyzed:
we found that they are able to reproduce colors, excitation and ionization
equilibria as well as the wings of H$_{\alpha}$. The comparison solar
abundances were carefully determined. From a new analysis of weak
Fe~I lines with accurate $gf$s (Bard \& Kock 1994) we derived
$\log \epsilon(Fe)_\sun=7.52$, in agreement with the Fe abundances
determined from meteorites and Fe~II lines.

We derived abundances for 21 species. Main results are:
\begin{itemize}
\item The metal abundances of the program stars span the range
 $-2.50<$[Fe/H]$<+0.17$.
\item Lines of most elements are found to form in LTE conditions. Fe lines
 satisfy very well the excitation and ionization equilibria. A comparison with
 statistical equilibrium computations shows that rather large collisional
 cross sections are required to reproduce observations. If these cross sections
 are then used in the analysis of the formation of Fe lines in subdwarfs and
 RGB stars, no significant departures from LTE are found for these stars,
 thus validating the very numerous LTE analyses.
\item RR~Lyraes share the typical abundance pattern of other stars of similar
 \feh : $\alpha$-elements are overabundant by $\sim$0.4~dex and Mn is
 underabundant by $\sim$0.6~dex in stars with \feh$< -1$. Solar scaled
 abundances are found for most of the other species, except for the low Ba
 abundance in the extremely metal-poor star X~Ari (\feh$\sim-2.5$).
\item Significant departures from LTE are found for a few species: Nd~II,
Ce~II,
 Y~II and Sc~II are severely underabundant ($\sim$0.5~dex) in metal-rich
 variables; Ti~I and Cr~I are slightly ($\sim 0.1 - 0.2$~dex) underabundant
 in metal-poor stars. These effects are attributed to overionization. We
 suggest that the photoionization of the alkaline earth-like ions is due
 to Lyman
 lines emission produced by the shock waves that propagate in the atmosphere of
 these variables (Fokin 1992).
\item Departures from LTE were considered in detail in the derivation of
 abundances for the light elements (O and Na). Significant corrections were
 required for the O~I IR triplet and the Na~D lines. The resulting pattern
 reproduces that observed in less evolved field stars. We did not find any
 evidence for an O-Na anti-correlation among these field HB-stars, suggesting
 that the environment is likely to be responsible for the
 anti-correlation found in metal-poor globular cluster stars (Sneden et al
 1992).
\end{itemize}

We used our new \feh abundances, as well as values from Butler and coworkers
(corrected to our system), and from high resolution spectroscopy of globular
clusters giants, to obtain a revised calibration of the low-resolution
metallicity index \DS (Preston 1959):
$${\rm [Fe/H]}= -0.194(\pm 0.011)\Delta S - 0.08 (\pm 0.18) $$
Our new metallicity scale is stretched on both low and high metallicity ends
with respect to Butler's (1975). The error in [Fe/H] by \DS observations
is 0.16~dex, well of the same order of high resolution metallicity
determinations. The slope of the calibration obtained considering only stars
with $4<$\DS$<10$\ is slightly smaller than that obtained using all stars.
While this difference is only barely significant, it might point out
the presence
of a non-linearity of the \DS {\it vs} \feh relation, as suggested by Manduca
(1981).

The new \feh values were used to update the metallicity calibration of the
\CaII index (Clementini et al 1991). Using the present new metallicities, and
$W'(K)$ values and relative errors from Clementini et al (1991), a
least-squares fit weighted both in $W'(K)$ and \feh gives:
$$ {\rm [Fe/H]} = 0.65(\pm 0.17)W'(K) - 3.49(\pm 0.39)$$

Finally, our new metallicity scale was used to revise the metallicity
dependence
of the absolute magnitude of \RR stars, M$_V$. Using M$_V$ values from
Fernley (1994) for the field stars, and estimates from Liu \& Janes (1990b) and
Storm et al (1994) for the cluster variables, we found:
$${\rm M}_V = 0.20(\pm 0.03){\rm \feh} + 1.06(\pm 0.04)$$
and:
$${\rm M}_V = 0.19(\pm 0.03){\rm \feh} + 0.96(\pm 0.04)$$
the last being obtained by using M$_V$ estimates derived for a value of the
conversion factor between observed and true pulsation velocity $p=1.38$\
(Fernley 1994). The adoption of the new metallicity scale does not yield
significant changes in the slope and zero-point of the M$_V$ {\it vs} \feh
relation. Observations do not rule out the possibility that the slope of the
M$_V$ {\it vs} \feh relation might be different for metal-poor and metal-rich
variables. However, a larger sample of Baade-Wesselink M$_V$\ determinations
is requested to definitely settle this question.


\end{abstract}

\newpage

\section{INTRODUCTION}

\noindent
This is the second in a series of papers dealing with abundance analysis of RR
Lyrae stars from high resolution spectroscopy (Clementini et al 1994a,
hereafter Paper~I). In Paper~I we successfully applied our technique to derive
abundances of three $ab-$type RR~Lyrae variables in the globular cluster M4.
Here we extend our study to a larger sample of bright field {\it ab}-type RR
Lyraes, using high quality observational material explicitly taken for this
purpose.

A number of factors make elemental abundance analysis of RR~Lyrae stars an
extremely worthwhile challenge:
\begin{itemize}
\item{(a)} RR~Lyrae variables are powerful tools to study the chemical
composition of the halo and disk of our Galaxy, and of the globular clusters.
RR~Lyraes have been used to derive the metal abundance of: (i) globular
clusters (Smith \& Butler 1978, Smith 1984, Costar \& Smith 1988), (ii) Baade's
Window (Walker \& Terndrup 1991), (iii) the Galactic Halo, and to study  the
metallicity distribution as a function of the galactocentric distance (Suntzeff
et al 1991, hereinafter SKK91). Since these variables are generally too distant
to allow a direct measure of their metallicity with high resolution
spectroscopy, the most commonly used method to derive their metal abundance is
via the spectrophotometric index \DS (Preston 1959). \DS measures the
difference in tenths of spectral class between the spectral type of an
$ab-$type RR~Lyrae at minimum light estimated from the hydrogen lines,
and that estimated from the \CaII intensity, and it is correlated to the \feh
abundance of the star. Butler (1975) empirically derived the following
metallicity calibration for $\Delta S$:
$$ [{\rm Fe/H}] = -0.16\Delta S - 0.23 ~~~(1),$$
using metal abundances measured from curve of growth analysis of 13 field RR
Lyrae stars. (We adopt the usual spectroscopic notation, namely: $[X] \equiv
\log (X)_{\rm star}-\log (X)_\odot$\ for any abundance quantity $X$, and
$\log \varepsilon (X)=\log (N_X/N_H)+12.0$\ for absolute number density
abundances). Butler's calibration was further confirmed by
the abundance analysis
of Butler \& Deming (1979). More recently Clementini et al (1991) have used the
equivalent width of the Ca~II K line, W$'$(K), to derive the metal abundance of
field RR~Lyraes. The W$'$(K)-[Fe/H] relation ([Fe/H]=0.53W$'$(K)$-$3.08) is
tighter than the one involving the \DS index;
however, it is again based on the Butler (1975) and Butler \& Deming (1979)
abundances. Butler's abundances,
however, are now rather old and his work ``needs to be redone with high S/N,
high dispersion digital data, using modern synthetic spectral codes" (Suntzeff
et al 1994, hereinafter SKK94).  In spite of the large use of RR~Lyraes as
metallicity indicators, the literature of the last decade contains
no modern redeterminations of the original metal abundances
on which the \DS and Ca II K line indices are based. In this
paper we present abundance analysis for 10 field RR~Lyraes, 4 of which taken
from Butler original sample, using high S/N ($>$200), moderately high
resolution ($\sim$ 18000) CCD spectra obtained with the Cassegrain echelle
spectrograph of the Palomar 60-inch telescope (McCarthy, 1988).
Our \feh abundances are used to derive a revised  calibration of the \DS and
\CaII indices. The new metallicity scale is then compared with the globular
clusters metallicity scale.

\item{(b)} \RR stars are primary distance indicators for
our own and nearby galaxies
because they are easily detected even at a large distance and exhibit a
relatively small dispersion in their intrinsic luminosities. RR~Lyraes have
been detected in the Magellanic Clouds (see e.g. Walker 1991), in M31 (Pritchet
\& Van den Berg 1987), and in a few galaxies of the Local Group (Saha et al
1992a,b and references therein). The absolute magnitude of RR~Lyraes (M$_V$)
is derived
with a rather well established accuracy by means of the Baade-Wesselink (B-W)
method (Liu \& Janes 1990a,b, Jones et al 1992, Cacciari et al 1992), which
also defines the \Mrr {\it vs} \feh dependence. We have used our \feh
abundances to revise the \Mrr {\it vs} \feh relation.

\item{(c)} RR~Lyraes can be used to investigate the origin and
evolution of the abundance anomalies found for giants in globular clusters.
In a series of papers, Kraft
and coworkers have found that in halo giants belonging to the field and to
globular clusters there is a global anticorrelation of [Na/Fe] and [O/Fe] (see
e.g. Sneden et al 1994). It is not clear if this anticorrelation extends to
metal-rich clusters ([Fe/H]$>-1$), since it has not been found for giants in 47
Tuc (Brown \& Wallerstein 1992; Carretta \& Gratton 1992) and M71 (Sneden et
al 1994). The most plausible interpretation for this anticorrelation is that
the O decline is the result of deep mixing, in which some $^{23}$Na produced by
proton captures on $^{22}$Ne is dredged up to the surface (Denisenkov \&
Denisenkova 1990; Langer et al 1993). While the effect is primarily correlated
to the evolutionary state, it seems to be modulated by some other mechanism,
which may be meridional circulation activated by core rotation (Sweigart \&
Mengel 1979). Atmospheric effects or departures from LTE are less likely to be
important (Drake et al 1992, 1993). Observations of Na and O abundances in RR
Lyrae stars may play an important role, since these stars are in an
evolutionary phase following that observed on the giant branch.
It should then be possible to observe some RR~Lyrae variables which have low
O and high Na abundances: this should
be explored by means of accurate spectroscopic analyses,
taking into account the possibility of departures from LTE.

\item{(d)} Quantitative estimates of statistical equilibrium in late type stars
have been up to now hampered by our poor knowledge of collisional cross
sections, mainly those with neutral H atoms; these last can only be obtained
theoretically, owing to difficulties in the related experiments. Steenbock \&
Holweger (1984) suggested that this mechanism plays an important thermalizing
r\^ole in late type stars, on the basis of order-of-magnitude estimates by
Drawin
(1968, 1969). Drawin formulas after Steenbock \& Holweger have since been
used in various estimates of non-LTE effects in late type stars (see e.g.
Steenbock 1985). However, recently Caccin et al (1993) used more reliable
estimates by Kaulakys (1985, 1986) to show that at least for Na, Drawin
formulas severely overestimate the cross sections for HI collisions. Given the
large theoretical uncertainties, a parametric approach may be used, where
collisional cross sections are estimated by matching observed non-LTE features
in stars where they are expected to be large. These cross sections may then be
used to predict departures from LTE in other stars, where they are expected to
be smaller. Due to the combination of atmospheric parameters (gravity $g$,
effective temperature \teff\ and overall metal abundance), and to the presence
of shock waves in their atmospheres caused by the pulsation mechanism itself,
departures from LTE are expected to be non-negligible in \RR stars. Since the
atmospheric parameters for abundance analysis can be determined from the light
and radial velocity curves of these variables without any {\it a priori}
assumption about excitation and ionization equilibria, any observed difference
in abundances derived from neutral and ionized species or among lines of
different excitation can be considered, reliably enough, as a hint of
departures
from the classical LTE assumption. Note, however, that this guess can be
misguided by uncertainties existing in the knowledge of the temperature
stratification for real variable stars compared to the hydrostatic equilibrium
model atmospheres adopted in the present analysis. This effect may be
significant owing to the dynamical character of \RR atmospheres. Hence, an
estimate of the reliability of the adopted model atmosphere must be performed
with appropriate tests, such as the ability to predict
colors or the profiles of strong lines. Anyway, if we are able
to obtain a quantitative estimate of
these effects on the derived abundances, we could also put a firm upper limit
on the influence of departures from LTE in the derivation of the chemical
composition for stars cooler and less luminous than \RR stars,
e.g. red giant branch (RGB) stars and subdwarfs.
\end{itemize}

The observational material is presented in Section~2, where we describe the
star selection, the data reduction procedures, the equivalent width
measurements, and the radial velocities determined from the spectra. The
adopted atmospheric parameters are discussed in Section~3,
where emphasis is given to
the determination of accurate temperatures using a new calibration of color
indices, and to a discussion on the ability of the new model atmospheres by
Kurucz (1992) to reproduce the spectra of RR~Lyraes at minimum light. The
results of the abundance analysis are given in Section~4; care was devoted to
a comparison with solar data, and statistical equilibrium computations were
performed for Fe, Na, and O lines. Results for all elements are also compared
with those obtained for less evolved stars. Section~5 presents new calibrations
of the \DS and \CaII indices; a comparison of our results with those obtained
from high dispersion spectroscopy of red giants in globular clusters; and
finally a discussion of the impact of our abundances in the M$_V-$[Fe/H]
relation for RR~Lyraes. Conclusions are summarized in Section~6.

\section{ OBSERVATIONAL MATERIAL }

\subsection{ Star selection and observations }

\noindent
Spectra of the program stars were obtained with the Cassegrain echelle
spectrograph of the Palomar 60-inch telescope during the four nights: 28~July
- 1~August~1993. The P60 echelle spectrograph (McCarthy 1988) was
operated in echelle grating mode using a 52.65~lines/mm echelle grating and
quartz prism cross-dispersers which yield a resolution R=$\lambda/
\Delta \lambda$=38000 per pixel, and large wavelength coverage
(3400~\AA$\leq\lambda\leq$9900~\AA).
Spectra are recorded on a TI 800$\times$800 pixels backside illuminated CCD
(15 $\mu$m pixel size). The slit width was set at 1.43 arcsec,
which projects to 2.1 pixels on the detector.

The FWHM measured from Th-Ar lines is $\sim$0.23~\AA\ at 4300~\AA. The echelle
grating of the P60 spectrograph has an off-plane angle
($\gamma$) of about 10~degrees
which produces a variable tilting along each order. In the data reduction phase
we have neglected this effect in order to simplify the reduction procedure.
This results in a mean degradation of the resolution of about 10\% for stellar
sources. However, in our spectra the resolution is limited to about 10000 by
the large amplitude velocity fields in the atmosphere of the program variable
stars. The data cover the spectral range 3400-9900~\AA\ with 65 orders,
partially overlapping for $\lambda\le 7000$~\AA.  Exposures of a Th-Ar
lamp were taken with the telescope in
the same position of the object to  perform the wavelength calibration. Flat
field and bias exposures were taken routinely at the beginning and end of each
night, particularly to monitor slight temperature fluctuations of the CCD
Dewar.

We took 50 spectra  of 10 ab-type \RR stars close to the minimum light of the
variables. For one of the program star, (namely SW And), we obtained also
spectra around the maximum light phase (see Section 4.1.5).
Spectra of 5 candidate red horizontal branch stars (RHB) were also
taken for the purpose of making a comparative abundance analysis. Here we will
confine our study to the \RR stars; observations  are in progress on a larger
sample of RHB stars using the 1.8 m telescope of the Asiago Astrophysical
Observatory. A hot bright star was observed at the beginning of each night and
used during the reduction phase to locate the echelle orders within the frame.

Since one of the primary purposes of our study was to obtain a new calibration
of the \DS and \CaII indices we included in our program some of the objects in
the Butler (1975)  and Clementini et al (1991) original lists. We ensured also
that our program stars had known \DS and covered a large range in metallicity
($-2.2<$[Fe/H]$<0.1$).
The sample of objects ultimately observed by us at Palomar had
4 stars in common with a sample of bright field RR Lyrae variables
being studied spectroscopically by Heath et al (1995).
The objects in common should
allow the two sets of results to be integrated once both independent
abundance analyses are complete.

To make an accurate abundance analysis we need knowledge
of the atmospheric parameters (gravity and effective temperature) of the stars.
The gravity can be derived from the radial velocity curves of the stars while
the effective temperature can be estimated from their color curves by means of
a model atmosphere (see Paper~I and Section~3 for a more detailed discussion of
these topics). In our selection we therefore chose objects for which recent
$BVRI$\ and possibly $K$\ light curves and accurate radial velocities were
available in the literature (see references in Sections 3.2 and 3.4.1).

The currently available model atmospheres (Kurucz 1992; hereinafter K92) are
stationary LTE models; however, due to pulsation, the use of a stationary model
atmosphere for an \RR is an approximation not always valid. It has been found,
in fact, that shock waves propagate throughout the atmospheres of these
variables (Preston \& Paczynski 1964, Gillet \& Crowe 1988, Clementini et al
1994b). The presence of shocks is associated with the appearance of the bumps
and humps in the light curve of the variables. Static model atmospheres are a
good approximation for these stars only if the regions of the shocks are
avoided. To meet these requirements, our
exposures were generally shorter than 60 minutes and our
spectra were all acquired near minimum light (see discussion in
Section~3.4). Since most of our targets have updated ephemerides, we could
safely locate the minimum light and properly phase our exposures. The list of
objects is shown in Table~1 together with the adopted ephemerides for
each object (Columns~4 and 5), the Heliocentric Julian Day (HJD) at half
exposure of each spectrum (Column~6), and the corresponding phase (Column~7).
Finally, the last column gives the references for the adopted ephemerides.

\subsection{Data reduction}

\noindent
The first part of the data reduction was performed using the facility ECHELLE
in the IRAF \footnote{ IRAF is distributed by the National Optical Astronomy
Observatories, which is operated by the Association of Universities for
Research in Astronomy, Inc., under cooperative agreement with the National
Science Foundation.} package.

We used spectra of four bright stars (one for each night) to locate the echelle
orders within the frame. The position of each order was then traced
interactively with a cubic spline. Assigning the bright star as a reference,
order finding and tracing was performed on science frames. Scattered light
(and, hence, also the bias) was then eliminated with a two-dimensional fit of
the interorder regions across and along the dispersion axis. The spectra
corresponding to the recorded echelle orders were extracted from each of these
cleaned frames.

Two-dimensional dispersion solutions were found for the Th-Ar arc spectra. The
typical r.m.s. deviation of the arc lines from the fitted wavelength
calibration polynomial was $\sim 0.02$~\AA. This can be regarded as the
internal accuracy in our wavelength calibration. The reduction of the arc
spectra confirmed the spectrograph stability during multiple observations of
the same object. Out of 65 orders found, we retained for further analysis only
those 52 that cover the spectral region between 3790 and 9160~\AA, because S/N
is low outside this range. Further reduction and data analysis were carried out
using the ISA (Gratton 1988) package, purposely designed for one dimensional
high resolution spectra. We found that division by lamp flats did not yield
satisfactory results at long wavelengths due to the presence of rather strong
interference fringes which were not well divided out with this technique.
Better results were achieved by dividing the spectra by a pseudo-flat field
obtained from the spectrum of X~Ari, the most metal-poor star in our sample,
for which a fiducial continuum may be easily identified by means of cubic
spline interpolation through selected spectral points. This technique allowed
us to properly correct interference fringes up to $\lambda <$7300~\AA. Beyond
this limit the only spectral feature we examined was the permitted O~I triplet
at 7771-7774~\AA\ (see Section~4.5). Single spectra of each star were compared
to eliminate the (very few) cosmic ray spikes, and then summed. Spectra were
not shifted before adding them together to account for differences in radial
velocity because the effect of the variation in radial velocity at the observed
phases is negligible at our spectral resolution. Giving the brightness of the
program stars and the small size of the slit, sky contamination can be safely
neglected: therefore no sky subtraction was performed on the spectra. Whenever
relevant (5860-5920, 6270-6320 and 7080-7130~\AA), telluric lines were divided
out still using the spectrum of X~Ari; however, regions including strong
telluric bands were not used in the analysis. In the case of the D lines
region, particular care was taken to prevent the stellar and interstellar
absorption lines present in the spectrum of X~Ari from producing false emission
features on the spectra of the other stars. For this purpose, we used also the
spectra of other metal-poor stars, having different radial velocities. Finally,
we traced the continuum on the co-added spectra by using a cubic spline
interpolation through a few selected spectral points. Comparison with synthetic
spectra showed that the final fiducial continuum was correct within 0.5~\%, at
wavelengths longer than 4500~\AA, with the exception of the region of the
Balmer lines where errors may be as large as 1-2\%. In Table~2 we list total
exposures and S/N of the co-added spectrum for each star. Small portions of the
co-added normalized spectra of X~Ari, RR~Cet and SW~And are shown in Figure~1.

\subsection{ Equivalent widths }

\noindent
Equivalent widths ($EW$s) for $\sim 100$\ lines in the average spectrum of
each star were measured by means of a gaussian fitting routine. This procedure
works quite well for RR~Lyrae stars because damping wings are weak for all
measured lines; in a few cases, comparisons with synthetic spectra were also
done. The complete list of
lines and adopted gf values and their measured
$EW$s is given in Table~3a,b,
and is available on electronic form from R.~Gratton. Errors in
these $EW$s are mainly due to uncertainties
in the location of the fiducial continuum: for this reason, $EW$s were not
measured at wavelengths shortward of 4500~\AA\ in the line rich spectra of
SW~And and V445~Oph, where continuum location is likely to be underestimated.
Errors in the $EW$s are not easy to accurately estimate because the measured
$EW$s should depend on phase and $S/N$\ level. A rather realistic estimate can
be obtained by comparing the $EW$s measured for the same lines in the spectra
of RR~Lyr and RR~Cet, which have similar atmospheric parameters at the average
observed phases (RR~Cet has only a slightly larger value of the microturbulent
velocity). This comparison is shown in Figure~2, where we have overimposed the
best-fit regression line with zero constant value: $EW_{\rm RR~Cet}=1.052(\pm
0.008)~EW_{\rm RR~Lyr}$\ from 76 lines; the regression coefficient is slightly
different from 1, mainly due to the different values of the microturbulent
velocity.
The scatter around this best-fit line is 9.3~m\AA, if we attribute equal errors
to the measure of $EW$s in each star, we obtain an r.m.s. error of 6.6~m\AA.
Figure~2 suggests that this error is fairly independent of the $EW$\ itself:
hence, 10~m\AA\ is a reasonable lower limit for reliable estimates of the
$EW$s. RR~Cet and RR~Lyr spectra have the highest $S/N$; errors in $EW$s,
$\sigma(EW)$, for the other stars likely scale as the inverse of $S/N$:
$\sigma(EW)\sim 2400/(S/N)$~m\AA. The values of $\sigma(EW)$\ for each star are
given in Column~6 of Table~2.

No direct comparison of our $EW$s with those measured by Butler (1975) and
Butler \& Deming (1979) is possible, since the spectra used in
those analyses were taken close to maximum light.

Carney \& Jones (1983) analyzed photographic echelle spectra of VY~Ser; their
mean phase of observation is 0.615, compared to our mean value of 0.72. On our
scales (see below), \teff\ and gravity at Carney \& Jones phase were
\teff=5969~K and $\log g=$2.60 (their adopted values are \teff=6000~K and $\log
g=2.3$), compared to the mean values for our observations of \teff=5993~K and
$\log g=$2.69. These sets of values are very similar,
so that no important modifications of the $EW$s are expected. A
line-to-line comparison of the $EW$s indicates that our values are smaller by
$7\pm 4$~m\AA\ ($\sigma$=16~m\AA: 15 lines). This difference is small, and a
correction of this entity would not affect significantly our analysis; however,
we can expect that our $EW$s are more reliable since our
CCD echelle spectra have much higher $S/N$.

\subsection{ Radial velocities and phases }

\noindent
Radial velocities were measured from the average spectra using a rather large
number of lines ($\sim 100$) for each star. They are listed in Column~4 of
Table~2. These values can be used to check the phases of our spectra. Since no
radial velocity standard was observed, the zero-point of these radial
velocities was determined using the telluric [OI] line at 5577.341~\AA.
Internal errors of these radial velocities, as deduced from the line-to-line
scatter (reduced by the square root of the number of lines), are $\pm
0.3$~\kms; star-to-star variations in the radial velocity for the telluric line
are similar ($\sigma=0.2$~\kms), showing that spectrograph flexures are small.
However, uncertainties in our radial velocities are larger, since errors in the
determination of the radial velocity from individual stellar lines are as large
as  2.7~\kms, where similar contributions are due to photon noise, residual
blending effects, and errors in the wavelength calibration; the last source of
error (amounting to $\sim 0.02$~\AA, i.e. $\sim 1.1$~\kms) is systematic, and
then apply also to the telluric [OI] line. On the other hand, we think that our
zero-point error is not very large, as confirmed by the comparison with the
expected radial velocities at the observed phases (Column~5 of Table~2) for
those stars for which they may be considered reliable; there are 7 such
variables (no radial velocity curve is available for VX~Her; and ST~Boo and
RR~Lyr radial velocities are more uncertain because these stars are affected by
the Blazhko effect). The mean difference between observed and expected
velocities is $V_r(obs)-V_r(exp)=-0.4\pm 0.9~$\kms. This test confirms that our
ephemerides are reliable. We think that the Blazhko effect, more than errors in
the adopted ephemerides, might be responsible for the higher discrepancy found
for RR~Lyr and ST~Boo. We discuss in Section~3.4 the effect on the atmospheric
parameters adopted for ST~Boo and RR~Lyr if phases inferred from the radial
velocity measures are used in place of phases derived from ephemerides.

Finally, our spectra can be used to test the presence of systematic variations
with optical depth, in radial velocities at minimum phase. This problem is of
some relevance in the determination of the absolute magnitude of \RR
variables with the B-W method, where the assumption is made that no
velocity gradients exist in the region of formation of the lines used to
measure the radial velocity (basically weak metal lines), and between this
region and the continuum forming region. Jones (1987) and Clementini et al
(1994b) measured the velocity of lines at different depths of formation in a
sample of field and cluster RR~Lyraes and reached the conclusion that no
significant velocity gradients ($\leq$ 2 \kms) exist throughout the atmospheric
layers where these lines are formed. Here we can repeat their test in a more
stringent way. Since the stronger lines measured on our spectra (with $EW\sim
300$~m\AA) form at much shallow optical depths ($\log \tau_{\rm Ross}\sim -3$)
than very weak lines (forming at $\log \tau_{\rm Ross}\sim -1$) \footnote{ We
verified this assertion by computing the average depth of formation (at line
center) of lines of different strength and excitation following the precepts of
Magain (1986).}, we have checked whether a correlation exists between $EW$\ and
$V_{\rm rad}$. The mean value for the linear regression coefficient $a$\ in the
$EW-V_{\rm rad}$\ plane over all the stars was $a=-2.3\pm 1.9$~\kms/\AA\ (where
the error bar is the standard deviation of the mean over all 10 program stars).
For comparison, in the same observing run we obtained spectra for a sample of
non-variable metal-poor giants with $4600\leq$\teff $\leq 6000$~K: the average
value of $a$\ for this sample was $a=0.2\pm 1.4$~\kms/\AA. The trend of radial
velocity with $EW$s measured in \RR variables is only barely significant;
we consider it as a marginal indication of a small acceleration of the outward
motion throughout the atmosphere at the observed phase (i.e. at minimum).

\section{ ATMOSPHERIC PARAMETERS }

\noindent
As we have anticipated in Section~2.1, to perform abundance analysis we need
the effective temperature, the surface gravity, the overall metal abundance and
the microturbulent velocity of our program stars. Effective temperatures
corresponding to the phases of our spectra were derived from the $B-V$, $V-R$,
$V-I$, and $V-K$\ colors of the stars using K92 models of appropriate
metallicity and gravity. Colors have been previously corrected for reddening.
The procedure used to estimate the reddening is described in detail in
Section~3.1. In Sections~3.2 and 3.3 we discuss the derivation of surface
gravity and overall metal abundance. In Section~3.4 we describe our
procedure to derive effective temperatures from K92 model atmospheres.
Finally, in Section 3.5 we discuss the adopted microturbulent velocities.

\subsection{ Reddening}

\noindent
Reddening is a crucial point in our analysis since an error of 0.01~mag in the
adopted reddening translates into an error of $\sim$50~K in the derived
temperature; this, in turn, translates into an uncertainty in the derived
abundance that for instance in the case of Fe~I corresponds to $\sim$0.05~dex.
Reddening estimates for the program stars based on photometric indicators have
been published by several authors (Sturch 1966, Jones 1973, Lub 1977a,b, 1979,
Liu \& Janes 1990a, Blanco 1992 and reference therein). Blanco (1992)
(hereinafter B92) presents a critical re-evaluation of the reddening of {\it
ab}-type \RR stars as derived from photometric indices. He uses a revised
version of Sturch (1966) method to evaluate $E(B-V)$\ from the observed
near-minimum light colors for {\it ab}-type RR~Lyraes, and derives a formula
which gives $E(B-V)$\ as a function of the metallicity of the star as inferred
from the \DS index, the period, and the $B-V$\ color during the phase interval
0.5$<\phi<$0.8:
$$E(B-V)=<B-V>_{0.5<\phi<0.8} + 0.0122\Delta S - 0.00045(\Delta S)^2-
0.185P-0.356~~~(2)$$
Equation (2) was established using: (i) mean colors at minimum light taken,
when available, from B-W analysis light curves (this is also the
photometry used in the present study to derive temperatures); and (ii) \DS
values collected from the literature and reduced to Butler's equivalent values.

B92 makes a very extensive comparison with previous reddening determinations,
particularly with those by Lub (1977a,b,1979), and Sturch (1966), which are
found to be in good agreement with his results within an accuracy of
$\pm$0.02~mag. A completely independent method to estimate reddening is
provided by Burstein \& Heiles (1978, 1982, hereinafter BH78 and BH82) who used
H~I column densities plus galaxy counts to determine the average reddening as a
function of Galactic latitude and longitude, and give values appropriate for
the RR~Lyraes (taking into account their distances). In Table~4 we
compare B92 reddening values for our stars with BH78 and BH82. Also listed are
reddening values used in the B-W analysis of the various stars. There is
generally good agreement between B92, BH78 and BH82 with the exception of
BH82 values for V445~Oph and RR~Lyr which deviate by a very large amount from
both Blanco's and BH78 estimates, and that therefore were not considered (see
also the discussion on the reddening of V445~Oph in Fernley et al 1990). In
conclusion, B92 reddening estimates were adopted for our stars with the
exception of V445~Oph for which a reddening value of 0.27~mag was found to be
more appropriate during the abundance analysis procedure; and of UU~Cet, for
which a lower metallicity compared to B92 was recently found with \DS
analysis by SKK94 (see discussion in Section~3.3). $E(B-V)$\ for this star was
derived from eq. (2),
taking into account the new metallicity estimate. The adopted reddening values
are listed in Column~2 of Table~5. We associate
with these estimates an uncertainty of $\pm$0.02~mag.

We may compare the adopted photometric estimates of reddening with those
derived from interstellar absorption features present in our spectra. We
measured the $EW$s for the Diffuse Interstellar Band (DIB) at 5780~\AA, and for
the interstellar Na~D lines. DIB $EW$s were transformed into reddening by means
of the calibration drawn by Herbig (1993), using only those stars (72) with
$E(B-V)\leq 0.4$, matching the expected range of reddening values for the
program \RR variables. The adopted calibration was $E(B-V)=(1.54\pm
0.10)~EW_{5780}$, where the $EW$\ of the DIB is in \AA. For the Na~D lines, we
transformed Na~I column densities deduced by means of the doublet ratio method
into reddening using the relation: $E(B-V)=(3.27\pm 0.19)~10^{-14}~n({\rm
Na~I})$, obtained by Benetti et al (1994) from a compilation of literature
data about interstellar lines and reddenings. Table~5 lists the
relevant data. The agreement between the different estimates of the
interstellar reddening is good for all stars, excluding X~Ari. The Na~D lines
give slightly larger reddening estimates, but the difference is not
significant. Mean differences are:
$$E(B-V)_{\rm Phot}-E(B-V)_{5780}=0.00\pm 0.02~{\rm mag}, \sigma=0.06~{\rm
mag}$$
$$E(B-V)_{\rm Phot}-E(B-V)_{\rm Na~D}=-0.03\pm 0.02~{\rm mag}, \sigma=0.04~{\rm
mag}$$
$$E(B-V)_{5780}-E(B-V)_{\rm Na~D}=-0.02\pm 0.04~{\rm mag}, \sigma=0.10~{\rm
mag}$$
If we average the two spectroscopic determinations, the mean difference with
the photometric reddening (again excluding X~Ari) is:
$$E(B-V)_{\rm Phot}-E(B-V)_{\rm Spec}=-0.01\pm 0.01~{\rm mag}, \sigma=0.03~
{\rm mag}$$
While this agreement may be fortuitous (given the large spread usually existing
in reddening estimates from these spectroscopic features), on the whole it
supports the reddening scale adopted in this paper.

Reddening in the $(V-R)$, $(V-I)$ and $(V-K)$ colors was obtained from
the $E(B-V)$\ values in Column 2 of
Table~5 using Cardelli et al (1989) absorption
coefficients: A($R$)/A($V$)=0.751, A($I$)/A($V$)=0.479 and A($K$)/A($V$)=0.114,
which are valid for a standard value of the total to selective absorption
R$_V$=A($V$)/$E(B-V$)=3.1. These coefficients are very similar to those
derived from Howarth (1983) formula. Cardelli et al (1989) coefficients are
calculated for the Johnson photometric system. The $V-R$ and $V-I$ colors used
in
the present paper are in the Cousins system. Transformation between the two
photometric systems is achieved using Bessell (1979) equations:
$(V-R)_C$=0.713$(V-R)_J$ and $(V-I)_C$=0.778$(V-I)_J$, that lead to the
following reddening corrections to apply to the observed colors:
$E(V-R)_C=0.550\,E(B-V)$, $E(V-I)_C=1.257\,E(B-V)$, and
$E(V-K)=2.747\,E(B-V)$. The
use of Howarth (1983) coefficients would give reddening corrections in good
agreement with those used here even for the most reddened stars, for which we
would find differences $\leq$0.02~mag.

\subsection{ Gravities}

\noindent
Gravity is an input parameter for abundance analysis and it is also needed to
select the appropriate K92 model to derive the color-temperature
transformations. Due to pulsation the gravity of an \RR star varies during the
cycle. The gravity to consider is therefore the {\it effective gravity} that is
described by the formula:
$$g=GM/R^2 + d^{2}R/d^{2}t ~~~(3)$$
where M and R are the mass and the radius of the star in solar units. The first
term in eq. (3) is the mean gravity of the star (i.e. the gravity that the
star would have if it were not pulsating), and may be derived from its mass and
mean radius; the second term represents the variation of the gravity along the
pulsation cycle due to the acceleration of the moving atmosphere and can be
estimated by differentiating the radial velocity curve of the variable.

Eight of the objects in our list have recently been the subject of B-W analysis
(Liu \& Janes 1990a, Cacciari et al 1989a,b, Cacciari et al 1992, Manduca et al
1981, Siegel 1982, Jones et al 1987, 1988, Fernley et al 1989, 1990). We have
used the masses and radii estimated for them with the B-W method to derive the
mean gravity term of eq. (3). Radial velocity curves are available for all
the variables in our list except VX~Her (Liu \& Janes 1989, Cacciari et al
1987, Clementini et al 1990, Sanford 1949, Carney \& Latham 1984, Jones et al
1987, Oke 1966, Preston \& Paczynski 1964, Clementini et al 1995), and have
been differentiated to derive the acceleration term in eq. (3). For
ST~Boo, which does not have $R$\ and $M$\ estimates, we have assumed a mean
gravity $\log GM/R^2$=2.79 and derived from eq. (3) $\log g$=2.71 as mean value
corresponding to our spectra, in good agreement with what found for stars of
similar metallicity. Lub (1977b) published $<\log g >$ values for \RR stars
estimated from Walraven photometric indices. He noticed that a zero-point error
may be present in his gravity calibration. Indeed we found that for the stars
we have in common our mean gravities are systematically lower than Lub's by
$-$0.28$\pm$0.09~dex. The gravity of VX~Her was then obtained from Lub (1977b)
converted to our gravity scale and further lowered by 0.07~dex to take into
account that our spectra are taken at minimum light. Mean effective gravities
corresponding to our average spectra are listed in Column 3 of
Table~10. Given the uncertainties in the adopted masses and radii
involved in the above procedure, we assign to our $\log g$ estimates
a conservative error of 0.20~dex.

\subsection{Metal abundance}

\noindent
In the literature we found a very large collection of metallicity estimates for
all the program stars. They include: \DS estimates (Preston 1959, Butler 1975,
McDonald 1979, Clube et al 1969, Alania 1973, Woolley \& Savage 1971, Smith
1990, Kinman \& Carretta 1991, SKK91, SKK94), \CaII estimates (Clementini et al
1991), and metallicities from abundance analysis by Butler (1975), Butler \&
Deming (1979) and Carney \& Jones (1983). B92 made an extensive review of the
various \DS values present in the literature which he reduced to a uniform
system and averaged for the stars with independent determinations. In
Table~6 we have collected the \feh values from:
\par\noindent
- Columns~2, 3 and 4: high resolution spectroscopy (Butler 1975; Butler \&
Deming 1979; Carney \& Jones 1983);
\par\noindent
- Column~5: \CaII index (Clementini et al 1991);
\par\noindent
- Columns~6 and 7: \DS parameter (SKK94; B92);
\par\noindent
- Column~8: the input values used in the B-W analysis;
\par\noindent
- Column~9: the input values used in the determination of \teff\ (see
below).

In general the  various estimates agree within 0.2~dex. UU~Cet and V445~Oph
deserve a more extended comment. SKK94 find for UU~Cet a rather low metal
abundance compared to other estimates. Since we do not have any reason for
preferring one estimate to the others, we have used the average among the
available values for this star: \feh=$-$1.2. As it will be discussed in
detail in Section~3.4, a metallicity \feh $\sim$+0.2 was found more appropriate
for V445~Oph during the abundance analysis.

With the exception of V445~Oph the values in Column~9 agree within
$\pm$0.10~dex with the mean of the other \feh abundances listed in
Table~6. A variation of 0.2~dex in the input metallicity has only a
marginal influence on the derived temperatures (see Section~3.4) and on the
derived abundances (see Section~4)\footnote{In fact, while we found that
important electron donors like Mg and Si are overabundant with respect to Fe in
the program stars, the effects of variations of the overall metal abundance by
a few tenths of a dex on the atmospheric structure and deduced abundance are
very small, because a large fraction of the free electrons is provided by H in
metal-poor \RR stars, as verified by appropriate computations.}.

\subsection{ K92 models and the color-temperature transformations}

\subsubsection{ Intrinsic colors for the program stars }

\noindent
Accurate $BVRI$\ photometry is available for all the stars in our sample
and $K$\ light curves are available for 6 of them. The observed
colors we used are on
the Johnson-Cousins system ($BV$\Rc\Ic\Kj). They are taken from the published
works of Liu \& Janes (1989), Fitch et al (1966), Siegel (1982), Manduca et al
(1981), Carney \& Latham (1984), Jones et al (1987, 1988), Fernley et al (1989,
1990), Burchi et al (1993), Clementini et al (1990, 1995), Cacciari et al
(1987, 1992), Barnes et al (1988), Stepien (1972), and Sturch (1966). Original
photometries were reduced to the Johnson-Cousins system using transformation
equations given by Bessel (1979, 1983) and Jones et al (1987). A detailed
description of the photometric data used in the present analysis and of the
procedure used to transform original photometries to a uniform photometric
system is given in the Appendix.

Data have then been re-phased according to the ephemerides in
Table~1, and smoothed curves were drawn through them. Colors read
from the smoothed color curves at phases corresponding to those of our spectra
were corrected for reddening according to the precepts given in Section~3.1.

\subsubsection{ The color-temperature transformations}

Effective temperatures corresponding to the phases of our spectra were
derived from the dereddened colors of the program stars using new color-\teff\
calibrations we explicitly determined for this purpose.
They were obtained using a procedure similar to that recently
followed by King (1993). This consists of two steps: first, an
empirical color-\teff\ calibration is determined for population~I dwarfs,
based on \teff\ derived by means of the Infrared Flux (IF) method (Blackwell
\& Shallis 1977). Second, the appropriate calibration for \RR stars is obtained
by correcting the calibration obtained for population~I dwarfs by the
corresponding offsets, due to differences in gravities and metal abundances,
given by K92 model atmospheres. The underlying assumption is that while
zero point errors may be present in the K92 theoretical colors, these
model atmospheres well predict the variations of colors with gravity and
metal abundances.

\medskip
\noindent
{\bf The color-\teff\ transformations for population~I dwarfs}

\medskip
\noindent
Fernley (1989, hereinafter F89) compared observed $V-K$\ colors and effective
temperatures $T_{\rm eff}$s derived mainly with the IF method
(Saxner \& Hammarb\"ack
1985, hereinafter SH85) of population I main sequence stars with  synthetic
$V-K$ and \teff\ derived from Kurucz (1979, hereinafter K79) models, and found
that theoretical and empirical $(V-K)$-$T_{\rm eff}$ relations have different
slopes. We have repeated Fernley's procedure on K92 models. An approach similar
to ours has recently been followed by King (1993); however, this last author
used the effective temperatures derived from the IF method (SH85) based on the
MARCS model atmosphere by Gustafsson et al (1975). A slightly different
temperature is derived using the IF method and K92 models. Blackwell \&
Lynas-Gray (1994; hereinafter BLG94) published effective temperatures derived
with the IF method and the new K92 model atmospheres for a large sample of
bright solar metallicity stars.  They also made a comparison with results from
IF method and K79 models and give in their Table~2 the appropriate temperature
corrections. Our procedure was the following: a list of Population I main
sequence stars of spectral type A-F-G, luminosity class IV-V and with well
determined effective temperatures was obtained by merging SH85 and BLG94 lists;
temperatures by SH85 were systematically corrected upward according to Table~2
of BLG94. The correction applied was on average $\sim$+52~K; BLG94 values were
adopted for the 8 stars in common.
The resulting sample of population I main sequence stars includes 57 objects
and is shown in Table~7.
$B-V$, $(V-R)_{C}$, $(V-I)_{C}$, and $(V-K)_{J}$ dereddened colors for these
stars were collected from F89, SH85, BLG94 and Cousins (1980) data sets; they
are shown in Columns~7, 8, 9 and 10 of Table~7, respectively.
Separate semi-empirical calibrations were obtained for each color by fitting
a second order polynomial to the
$(V-R)_{C}$, $(V-I)_{C}$, and $(V-K)_{J}$ -\teff\ data, and a third order
polynomial to the $(B-V)$-\teff\ data, in the temperature range
$5000<\ $\teff$<8300$~K. The polynomial fits are given below and shown in
Figure~3 (solid lines).\\
$$T_{\rm
eff}=-5718.4077(B-V)^3+10088.399(B-V)^2-9316.128(B-V)+9115.8314~~(4a)~~
(57~{\rm objects})$$\\
$$T_{\rm eff}=3350.38062(V-R)^2_C-9445.28906(V-R)_C+8757.94727~~(4b)~~(22
{}~{\rm objects})$$\\
$$T_{\rm eff}=753.906677(V-I)^2_C-4836.16016(V-I)_C+8801.4248~~(4c)~~(22
{}~{\rm objects})$$\\
$$T_{\rm eff}=300.313507(V-K)^2-2447.8403(V-K)+8768.1709~~(4d)~~(54
{}~{\rm objects})$$\\
Deviations from the polynomial fittings are $\leq \pm$0.02 mag in $V-R$ and
$V-I$, and $\leq \pm$ 0.05 mag in $B-V$ and $V-K$ (with 80\% of the objects
within $\pm$0.03 mag).

We have compared the coefficients of our polynomial fitting for the
$(V-K)$-\teff\ calibration with BLG94 (lower line of BLG94 Table~8). In the
temperature range 5700$<$\teff$<$6700 K the two calibrations agree within 10~K.
The discrepancy is of about 40~K at \teff=5100 and 7700~K, and of about 100~K
at \teff$\sim$9000~K, with our temperatures systematically cooler. Our
calibration is extrapolated beyond 8300~K, and hence not reliable. We estimate
that equations (4) can be used to derive reliable \teff\ from observed $B-V$,
$(V-R)_{C}$, $(V-I)_{C}$, and $(V-K)_{J}$ colors within the temperature
interval 5000$<$\teff$<$8000~K. The dashed lines in Figure~3 represent K92
theoretical color-temperature calibrations for the same metal abundance and
gravity of the dwarf sample in Table~7 (i.e. [A/H]=0.0 and $\log
g$=4.0). We find that while theoretical and semi-empirical $(V-K)$-\teff\
calibrations are almost indistinguishable (upper panel of Figure~3), in the
temperature range of our interest (i.e. around 6200~K), the $B-V$, $V-R$ and
$V-I$ theoretical calibrations lie systematically below their semi-empirical
counterparts.

\medskip
\noindent
{\bf The color-\teff\ transformations for \RR stars}

\medskip
\noindent
We have corrected K92 synthetic colors to tie them to the semi-empirical
calibrations on the assumption that the offset between synthetic and observed
color is independent of gravity and metallicity\footnote{A comparison
between the temperature calibrations of the $b-y$\ index using K92 and the new
OSMARCS models (Edvardsson et al 1993) actually suggests a quite different
metallicity dependence, likely due to a different treatment of convection
between these two {\it latest generation} model atmospheres.  Unfortunately,
we do not
have predictions of $V-K$\ colors with the OSMARCS models. However, the $V-K$
colors may be empirically calibrated using the temperatures by Edvardsson et
al (1993) and Nissen et al (1994; these last authors use the same procedure);
temperatures obtained by this procedure are similar to those obtained by the
K92 calibration for population I stars, but they are much lower (up to 250~K)
for the most metal-poor stars, yielding much lower [Fe/H] values (by up to
0.4~dex). This would exacerbate the discrepancy between abundances for globular
clusters derived from giants and RR~Lyraes (see Section~5.1.1).
We then decided to consistently use K92 models
(which are the only {\it latest generation} model atmospheres available for RR
Lyrae) for the color-\teff\ calibrations throughout this paper.}. We have
calculated the offsets in color $\Delta (B-V)$, $\Delta (V-R)$, and $\Delta
(V-I)$ between K92 synthetic colors at [A/H]=0.0 and $\log g$ =4.0, and K92
colors for the gravity and the metallicity required for each star, and we have
subtracted these offsets from the semi-empirical calibrations. We have thus
generated a new set of K92 synthetic color-temperature transformations tied to
the semi-empirical calibrations. From them we have read the temperatures
corresponding to our spectra.

In Table~8 we list the temperatures of the average spectra derived
from the various color indices. Temperatures for individual stars generally
agree with each other within 200~K with $V-K$ and $V-R$ usually giving the
cooler values. The scatter is much higher ($\ge$300~K) if we use \teff\ from
K92
models not corrected for the semi-empirical calibrations. The effective
temperature derived from the $(B-V)$\ color of VX~Her is very discrepant and
was not considered; VX~Her and RR~Lyr are the stars which have the oldest and
least accurate photometry among our objects. Further, we did not use \teff\
from $B-V$\ for X~Ari since the $B-V$\ color for this star is affected by the
emission associated with the bump in the light curve (Gillet \& Crowe 1988).
Incipient H$_{\alpha}$\ emission was clearly detected in our spectra. V445~Oph
exhibits a large discrepancy between temperatures derived from different
colors. This is the most reddened object in our sample and some uncertainty
exists on the reddening and on the appropriate absorption coefficient R$_{V}$
to use for this star. The metallicity of the star is controversial too (see the
discussion in Cacciari et al 1992). We have iteratively tried various
combinations of reddening, R$_{V}$ and metallicity and used in the end the
combination $E(B-V)=0.27$, R$_{V}$=3.1 and \feh=+0.2, which seems to give more
consistent temperatures if we do not consider the \teff\ derived from $V-K$. On
the other hand, if we compare the dereddened colors of V445~Oph and SW~And
(which according to their \DS should have similar metallicities), we see that
while $B-V$, $V-R$ and $V-I$ colors of the two stars in the phase range of our
spectra (i.e. 0.5$< \phi <$0.7) agree within $\pm$0.02 mag, the $V-K$\ color of
V445~Oph is 0.1~mag redder than SW~And one. On the basis of these
considerations we chose not to use the $V-K$\ temperature of V445~Oph. The
temperatures not used are given in parenthesis in Table~8.

Since the $V-K$ - temperature calibration is not affected by any discrepancy
between synthetic and observed colors and it is less metallicity dependent (see
F89 and references therein),  we adjusted the temperatures derived from $B-V$,
$V-R$ and $V-I$ to the $V-K$ one by correcting them for the systematic mean
differences between temperatures derived from individual colors and $V-K$
($-49$~K, $-$51~K and $-$97~K for the $B-V$, $V-R$ and $V-I$ colors,
respectively). Finally, average temperatures were calculated as the mean of
these corrected $T_{\rm eff}$s;
they are listed in Column~6 of Table~8
together with the related r.m.s. (Column~7). The typical internal uncertainty
is 25~K.

Besides the internal errors, uncertainties in the derived temperatures depend
on: (i) the assumed gravities and metal abundances; (ii) uncertainties in the
assumed reddening; (iii) the accuracy of the photometric data; and (iv) the
theoretical soundness of using static model atmospheres to reproduce the
dynamical atmospheres of pulsating variables (this last issue is discussed in
the next subsection). Uncertainties in the assumed gravity affect the
color-temperature transformations only marginally; in fact,
irrespective of metallicity and color index, a
variation in $\log g$ from 2.5 to 3.0 produces a
variation in the temperatures derived from K92 models generally $\leq$40 K (see
also Figure~3 of Liu \& Janes 1990). To enter K92 models we have assumed,
therefore, a constant
gravity $\log g$=2.75 as appropriate for all stars; this value corresponds to
the average of Column~3 in Table~10. The sensitivity of the
color-temperature transformations to the input metal abundance has been checked
for a metal-rich and a metal-poor object. Tests made on SW~And (\feh=$-0.10$)
and VY~Ser (\feh=$-1.75$) show that the effect is relevant only for
temperatures derived from $B-V$ and $V-R$, and only for metal-rich objects. In
fact, a variation of 0.3~dex in metallicity translates into a variation of the
estimated temperature from $B-V$ of $\sim$100~K, and from $V-R$
of $\sim$50~K, at \feh=$-$0.10, or
$\sim$ 30~K and $\sim$6~K respectively at \feh=$-$1.75. For
the other colors the effect is $\leq$10~K.
Recall that the reddening accuracy is
$\pm$0.02~mag in $E(B-V)$ (i.e. $\sim$100~K); we can safely assume that the
errors due to the colors are $\leq$0.01 mag (i.e. $\sim$50~K), since we read
colors from smoothed curves where the accuracy of single data points is
$\pm$0.01~mag. These contributions add in quadrature and lead to an error on
the derived \teff of $\sim$115~K. The adopted temperatures are listed in
Column~2 of Table~10.

As we have discussed in Section~2.4, RR~Lyr and ST~Boo exhibit the highest
discrepancy between phases estimated from ephemerides and phases derived from
radial velocities measured from the average spectra. We find that if these last
values are used, the adopted gravities should be lowered by 0.10-0.15~dex for
both stars and the adopted temperatures should be $\sim$ 40 K and 60~K cooler
for ST~Boo and RR~Lyr respectively; these values are within the uncertainties
adopted for these atmospheric parameters.

\medskip
\noindent
{\bf \teff's from H$_{\alpha}$\ profiles}

\medskip
\noindent
The procedure we have followed to derive effective temperatures from the
observed colors of the program stars, and the abundance analysis technique we
will describe in Section~4, are based on the assumption that K92 model
atmospheres are able to reproduce well the \RR variables at minimum light.
In the following we shall briefly discuss this assumption. Figure~3 shows that
some discrepancy exists between the observed and K92 synthetic colors for
dwarf stars. This problem is enhanced for \RR stars where differences
between temperatures derived from color calibrations obtained by K92 models not
corrected for the empirical calibration may rise to more than 300~K. The
discrepancy is reduced to less than 80~K (although not totally eliminated) by
correcting for the semi-empirical calibrations derived for dwarf stars.

An independent estimate of the temperature of the continuum forming region is
provided by the profile of the H$_{\alpha}$\ absorption line. Our Palomar
spectra are not well suited to this purpose, owing to the quite large
uncertainties in the location of the continuum level in this spectral region
due to the presence of strong diffraction fringes, and to the narrow free
spectral range. We therefore acquired additional spectra (resolution $\sim
15,000$\ and $S/N\sim 130$) for three of the program stars (SW~And, RR~Cet, and
X~Ari) using the REOSC Echelle spectrograph at the 182~cm Copernicus reflector
of Asiago Astrophysical Observatory. This spectrograph uses a 79~gr/mm echelle
grating (so that the free spectral range is 1.6 times that at Palomar) and a
large format front illuminated Thomson CCD detector, with no appreciable
diffraction fringes. Spectra of several subdwarfs were also acquired during the
same runs with the same instrumental set-up, but higher $S/N$\ ($>250$): some
of them were used here to further compare our \teff\ from H$_{\alpha}$\
profiles with those derived from our calibration of photometric indices, as
well as those obtained by Fuhrmann et al. (1994) also from H$_{\alpha}$\
profiles.

All Asiago spectra were reduced using a homogenous procedure. We find that
after flat fielding by means of the spectra of a quartz iodine lamp reflected
by the dome, a residual trend of the continuum with wavelength still existed.
This
trend repeated very regularly through different orders, so that it was
efficiently eliminated by using the same average pseudocontinuum (obtained by
automatic IRAF routines) for all orders. We attribute this effect to the
different way the slit is illuminated when using light from the star or
from the
lamp. The spectra were then wavelength calibrated using a Th lamp, and
telluric lines were divided out by a synthetic spectrum constructed using
wavelengths and equivalent widths from Moore et al. (1966)\footnote{ Line
intensities were adjusted to fit telluric lines detected in each individual
spectra.}.

H$_{\alpha}$\ profiles obtained by this procedure were then compared with
synthetic profiles computed using K92 models and our own spectral synthesis
code, which included Doppler, natural damping, resonance (Ali \& Griem 1965,
1966), and Stark (Vidal et al 1970) broadening, following prescriptions
similar to those adopted by Fuhrmann et al. (1993, 1994). These
prescriptions should be correct for electron densities above 10$^{12}$
electrons per cm$^3$\ (roughly corresponding to $\log \tau\sim -2$\ for the
\RR variables). However, line formation in the outer part of the atmospheres
is likely affected by appreciable deviations from LTE, as demonstrated by the
presence of some incipient emission at wavelengths slightly shorter than the
core of the lines: hence these part of the profiles (as well as those
contaminated by other lines) were not considered in the \teff derivations. A
comparison with the solar flux spectrum (Kurucz et al 1984) showed that our
computed profiles are a little bit too narrow, so that the solar \teff we
derived is 6017~K. We then systematically corrected downward our \teff's by
247~K. Following this procedure, our \teff's are relative to the Sun. We note
that the profiles used for the \RR variables were computed
assuming $\log g=2.75$\
and [A/H]=$-1$; however, this assumption is not critical since Fuhrmann et
al (1993) have shown that when opacity is dominated by H$^-$\ and most
electrons are provided by H, the H$_{\alpha}$\ profile is almost independent
of surface gravity and metal abundance.

A comparison of observed and computed profiles is given in Figure~4. Our
\teff's from H$_{\alpha}$\ profiles are listed in Table~9
both for the three RR~Lyrae variables and the seven subdwarfs.
In this table we also
give \teff's from our photometric calibrations and those deduced by Fuhrmann
et al (1994) from H$_{\alpha}$\ profiles. On average, our H$_{\alpha}$\
\teff's for the
dwarfs are slightly  larger than both those derived from colours and by
Fuhrmann et al; however
differences ($27\pm 41$~K and $32\pm 34$) are not significant.
The r.m.s. scatter
of the star-to-star residuals (108 and 89~K respectively) indicates
that the internal
errors are $\sim 100$~K; we attribute these errors to small ($\sim 1$\%)
uncertainties in the location of the continuum level. In the case of
the \RR variables, the average difference between \teff's from
H$_{\alpha}$\ profiles and
those derived from the photometric
indices is $-73\pm 75$~K: again this difference is not significant. The
scatter of the results for individual stars (132~K) is consistent with an error
bar similar to that obtained for dwarfs.

We conclude that the comparison with the $H_{\alpha}$\ profiles supports our
color-temperature calibrations, and confirms that K92 models are able to
reproduce the temperature stratification of the continuum forming region
($\log \tau >-2$) of an \RR at minimum light to within an accuracy of
$\sim 150$~K.

\medskip
\noindent
{\bf K92 models and the atmospheres of \RR ~stars at minimum light}

\medskip
\noindent
The suitability of K92 models to reproduce the regions of formation of the weak
metal lines used in the abundance analysis ($\log \tau \sim -1 \div -2$) for RR
Lyrae stars close to the minimum light is
demonstrated by their capability to well
reproduce the ionization and excitation equilibrium conditions for elements of
the iron group as well as for the $\alpha-$elements (see Section~4), with no
need for deviations from LTE conditions. Deviations from LTE, possibly induced
by shock waves propagating throughout the atmosphere of the RR~Lyraes are
perhaps required only to explain the anomalous behaviour of some peculiar
elements (see Section~4.4); however, only small amplitude shock waves are
required, which are not expected to alter the global atmospheric structure of
the star. Beyond these regions (i.e. at $\log \tau < -2$) we do not have good
tests available to check whether K92 model atmospheres are still appropriate;
however, the asymmetry exhibited by the core of the H$_{\alpha}$\ line, as well
as its large variations from one spectrum to the other, would suggest that
deviations from LTE take place in the outer atmospheric regions.

To conclude, as far as abundance or B-W analyses are concerned, i.e. as far as
the regions of interest are confined to $\log \tau =0 \div -2$, K92 models
seem adequate to reproduce the atmospheres of \RR stars at minimum light.
They certainly represent an improvement upon K79 model atmospheres which gave
excitation and ionization temperatures higher by about 250 K than
temperatures derived from colors. The comparison between K92 and K79 models is
shown in Figure~5.

\medskip
\noindent
\subsection{ Microturbulent velocities }

\medskip
\noindent
Microturbulent velocities \vt\ were determined by zeroing the slope of
abundances with $EW$s for individual Fe~I lines. Since $EW$s were deduced for
numerous lines on both the linear (which is very extended in the spectrum of
RR Lyrae variables) and flat parts of the curve of growth, the values of \vt\
can be accurately determined from our spectra: typical internal errors are
$\pm 0.2$~\kms.

An assumption inherent to these determinations of \vt\ is that
the microturbulent velocity is
constant throughout the stellar atmosphere. There is no sound argument
supporting this assumption apart simplicity; in fact, non-constant
microturbulent velocities have been proposed for the Sun (see e.g. Maltby et
al. 1986), and the presence of shock waves makes it likely that \vt\ is not
constant in the atmospheres of the RR Lyrae stars too.
In the following we explore more in detail the aftermath of
the assumption of constant \vt.

The run of \vt\ with optical depth is difficult to be determined since it
cannot be presently derived on a theoretical base alone. On the other side,
very high resolution, high S/N data are required to disentangle the
contributions of micro and macro-turbulence from line profiles. The only
data available for RR Lyrae variables are the $EW$s; we then tried to
estimate the range of possible variations of \vt\ within the atmosphere of a
typical RR Lyrae star at minimum light
from the trends of the abundances deduced from
individual lines with $EW$s, exploiting the fact that the depth of formation
of lines is a function of line strength. In fact, if \vt\ is not constant
throughout the atmosphere, we expect that significant trends may be present
in the run of abundances with line strength even though the average slope has
been set at zero by adopting a suitable constant value for \vt.

In Figure 6a we plotted the residuals of the abundances deduced from individual
Fe lines minus the average Fe abundances, against $EW$s (in order to improve
statistics, we considered all 10 program stars together). The scatter of
individual points in Figure 6a is rather large; however, the correlation of
these residuals with a quadratic least square relation is highly significant
(about 5 standard deviations), while as expected the linear correlation
coefficient is very close to zero; this means that significant trends are
indeed present in the run of abundances with $EW$s, even though the average
slope is zero. It should be noticed that deviations from zero are not large;
this is shown by Figure 6b, where we plotted the average values of residuals
computed in bins of 0.2~dex, against $EW$s: the largest deviations are
$<0.1$~dex. Similar small trends could well be due to causes other than
systematic variations of \vt\ with optical depth, like e.g. deviation of
observed line profiles from gaussians (used to derive the $EW$s), inappropriate
consideration of damping, numerical approximations in the analysis code, etc.
Hence, we do not think that the plots of Figures  6a and 6b conclusively show
that \vt\ changes with optical depth in the atmospheres of RR Lyrae variables
at minimum light.

Anyway, in order to estimate the impact of these possible changes on our
abundance derivations we must have at least an order of magnitude estimate of
their values. This was obtained by comparing the run of Figure 6b with those
given by a standard abundance analysis (with constant microturbulent velocity)
performed on $EW$s of synthetic lines forming in a model atmosphere having
\teff=6200~K, $\log g$=2.75, and [A/H]=$-1.3$\ (these values being assumed
as representative of an RR
Lyrae star at minimum light). Various \vt$(\tau)$\ laws, where
\vt\ varies quadratically with log optical depth, were considered when
computing synthetic profiles (quadratic laws were considered because they
resemble that adopted by Maltby et al 1986 for the Sun). We finally selected
for our comparison those cases in which the analysis of the synthetic lines
yields a (constant) value of \vt=4~\kms and an amplitude of the run of the
abundance residuals with $EW$s similar to that obtained in our analysis of RR
Lyrae stars at minimum. These runs obtained from synthetic lines are also
shown in Figure 6b. Qualitatively, we found that a suitable quadratic
\vt$(\tau)$\ law matches reasonably well the observations.
The correct amplitude
and the position of the minimum at $\log EW/\lambda\sim -4.6\div -4.7$\ are
reproduced by adopting a minimum value of \vt$\sim 0.6$~\kms at a very shallow
optical depth ($\log \tau\sim -3$), and a large variation of \vt with optical
depth (\vt $\sim 8$~\kms) at $\log \tau\sim 0$. In fact, if we assume that the
minimum value of \vt\ is deeper in the atmosphere, the observed amplitude of
abundance variations with $EW$s is reproduced by much smaller variations of
\vt\ with optical depth, but the minimum in the run of abundances with $EW$s
shifts at $\log EW/\lambda>-4.5$, in disagreement with data. A value of $\log
\tau\sim-3$\ for the location of the minimum of \vt\ is similar to that
assumed by Maltby et al., close to the temperature minimum in the solar
atmosphere. On the other hand, the value of \vt\ at minimum (which is required
in order to place the minimum in the run of abundances with $EW$s at
$\log EW/\lambda\sim -4.6\div -4.7$) seems quite low. This casts some doubts
over the whole procedure.

Even though not entirely convincing, we now have a guess for the \vt$(\tau)$\
law. We then run our abundance analysis for a test star assuming this
law of variation of \vt, and then compared these results with those obtained
assuming a constant \vt. The differences between the abundances obtained in
the two cases yield a rough estimate of the possible influence of the
assumption of a constant \vt\ on our final abundances. These differences are
given in Column 7 of Table~11; in most cases they are less than 0.1~dex.
Elements with few, very strong lines exhibit larger variations ($\sim
0.15\div 0.20$~dex), which are very sensitive to line strength (e.g.
variations have the opposite sign for Ba~II lines).

We conclude that present data do not warrant the adoption of a variable \vt\
with optical depth in our analysis; however, would \vt\ be indeed variable, we
do not expect variations in the deduced abundances larger than 0.1~dex, except
for those elements for which we measured $EW$s for only a few, strong lines;
in these cases, variations as large as $0.15\div 0.20$~dex may be expected.

\subsection{ Adopted atmospheric parameters }

\noindent
The adopted atmospheric parameters are summarized in Table~10.
Gravities, \teff\ and microturbulent velocities
are as discussed in Sections~3.2, 3.4 and 3.5.
Metallicities are
instead slightly different than the values discussed in Section~3.3. They were
obtained iteratively during the abundance analysis procedure and are close to
the final derived values of [Fe/H].

\section{ ABUNDANCE ANALYSIS }

\noindent
Abundances were derived from the $EW$s using model atmospheres extracted from
the grid by K92. Table~11 shows the sensitivity of the derived
abundances on the values adopted for the atmospheric parameters.
Values in Table~11 were
obtained by comparing the derived abundances for RR~Cet with those obtained by
varying the atmospheric parameters
one at a time within the estimated range of confidence. The main effects are
those due to \teff\ on the abundances drawn from neutral species, and to
$\log g$ on the abundances derived from singly ionized (dominating) species,
while the overall metal abundance and the microturbulent velocity
(which is typically very large in an RR~Lyrae variable compared with a
non-variable giant) play only a minor role. Sensitivity of element-to-element
abundance ratios on the adopted atmospheric parameters are then much reduced
(to $\leq 0.05$~dex) if neutral species are compared with other neutral species
(e.g. with Fe~I), and singly ionized species with other singly ionized species
(e.g. with Fe~II): this is the adopted procedure throughout this paper.
However, there are significant exceptions: the abundances derived from high
excitation lines of dominant species (O~I, Si~II, S~I) are sensitive to both
gravity and temperature (with a trend opposite to that of Fe I lines); the
abundance derived from a single, strong (resonance) line like that of Al~I is
very sensitive to the microturbulent velocity. Abundance ratios to Fe derived
from lines of these species have then larger uncertainties ($\sim
0.10-0.15$~dex).

Column 6 of Table~11 shows the variation of the solar reference
abundances (obtained using the same line list used for the program stars) when
the constant flux solar
model atmosphere by K92 is replaced by Holweger \& M\"uller (1974;
hereinafter HM) empirical model atmosphere, which is considered to be
the best representation of the solar photosphere. Although the values in
this column depend somewhat on the adopted line list, they provide a rough
estimate of how the uncertainties in the abundances are related to the
structure of the atmospheres of metal-rich non-variable stars. These values
are generally small ($<0.06$~dex), and testify the significant progress made
with the new generation of atmospheric models. For consistency, hereinafter
we will use the abundances obtained from Kurucz models.

\subsection{ Fe abundances }

\noindent
Table~12 lists Fe abundances derived for the program stars. As
mentioned in the Introduction, the determination of abundances for \RR
stars offers a unique chance to test empirically the reliability of abundances
determined from high dispersion analysis; furthermore, departures from LTE
might be large in the warm and rarified atmospheres of \RR variables. We
expect that inadequacies of the assumptions (LTE, structure of the model
atmospheres, etc.) will show up as trends of the derived abundances from
individual lines with excitation and ionization, as well as
element-to-element abundance ratios not fitting in the usual pattern observed
in other old stars. However, in order to derive significant estimates of these
inconsistencies, considerable care must be devoted to the derivation of an
accurate set of oscillator strengths $gf$s; some insight can be obtained by a
comparison with solar abundances. For this reason, results obtained for the Sun
with the same code and set of $gf$s are also given in Table~12.

\subsubsection{ The Solar Fe abundance }

The solar Fe abundances listed in Table~12 require some more comments;
in fact, in the last years there has been some debate about the determination
of the solar
photospheric Fe abundance, since the value obtained using precise $gf$s
from the absorption experiments of the Oxford group
combined with
laboratory determinations of the damping parameters and the HM model atmosphere
(log~$\epsilon$(Fe)=7.63: Simmons \& Blackwell 1982) is larger than those
obtained from Fe~II lines (log~$\epsilon$(Fe)=7.48: Holweger et al 1990;
log~$\epsilon$(Fe)=7.54: Bi\'emont et al 1991; log~$\epsilon$(Fe)=7.48:
Hannaford et al 1992) and from meteorites (log~$\epsilon$(Fe)=7.51: Anders \&
Grevesse 1990). New sets of $gf$s for Fe~I lines, based on accurate lifetimes
and branching ratio measurements, have been published by
Bard et al (1991),
O'Brian et al (1991), and more recently by Bard \& Kock (1994). These sets of
$gf$s agree very well with each other: mean difference over the 29 strongest
lines is $0.006\pm 0.008$~dex, O'Brian et al $gf$s being slightly smaller than
those by Bard and coworkers, with a standard deviation of 0.044~dex for
individual lines. The scatter is larger ($\sim 0.15$~dex) for weaker lines,
likely due to random errors in the branching ratios used by O'Brian et al
(since a solar analysis using this set of $gf$s for weak lines gives a large
scatter), but the overall agreement is still good. The solar Fe abundance
determined using $gf$s from Bard et al (1991) is much closer to the meteoritic
value (log~$\epsilon$(Fe)=7.47: Holweger et al 1991) than that derived by
Simmons \& Blackwell; however, lines used in that analysis are strong, most of
them are not clean from blends, and the derived abundances depend on
assumptions about damping, which are uncertain, and for which
an average value was derived
from the solar analysis itself. To have a better insight into this aspect, we
redetermined the photospheric Fe abundances using the $EW$s and $gf$s by
Simmons \& Blackwell (1982) and Holweger et al (1991), as well as lines with
$gf$s measured by Bard \& Kock (1994) and $EW$s from the list of clean lines in
the solar intensity spectrum by Rutten \& Van den Zalm (1984). This last group
consists of lines which are much weaker than those used by Holweger et al
(1991) and Simmons \& Blackwell, and hence less dependent on the assumption
about collisional damping (for which we
used a mean relation drawn from the Simmons \& Blackwell data when reanalysing
all three sets of Fe line data).
The mean Fe solar abundances provided by these three groups of lines
(i.e., Simmons \& Blackwell 1982; Holweger et al 1991; and Bard \& Kock 1994
with Rutten \& Van den Zalm 1984)
using our code and the K92 solar atmosphere with a depth-independent
microturbulent velocity of 0.9~\kms~ were log~$\epsilon$(Fe)=7.59, 7.50,
and 7.52 respectively. We consider the last of our redeterminations
(log~$\epsilon$(Fe)=7.52; which agrees very well with the meteoritic
value) to be the most reliable.
Therefore the following procedure was adopted in this paper: whenever
possible, $gf$s for the Fe~I lines detected in the spectra of our targets
were taken from papers of the Oxford group
(see, for references, Simmons \& Blackwell, 1982) and Bard et al (1991),
[we could not use Bard \& Kock (1994) because lines in their list are too
weak to be measured in the spectra of the RR Lyraes]; the
$gf$s from the Oxford group were corrected upward by 0.03~dex to account
for the systematic
difference with those by Bard et al.
This systematic difference might be attributed to a
small error in the zero-point of Oxford $gf$s, which anyhow must be ultimately
based on lifetime measurements. $gf$s for Fe~II lines were taken from Heise \&
Kock (1990), Bi\'emont et al (1991) and Hannaford et al (1992). For all
remaining lines (taken from the list by Rutten \& van der Zalm 1984, and
Blackwell et al 1980), solar $gf$s were derived using the HM model atmosphere
and the Fe abundance derived from the other lines. As shown in Table~12
the solar Fe abundance obtained with this set of $gf$s for Fe~I is slightly
larger than that derived from Fe~II lines; however, this difference is small
when using the K92 model and does not cause serious concern; the trend with
excitation potential is also very small in the Sun.

\subsubsection{ Trends with excitation and ionization }

On the whole, the observed trends for Fe abundances measured in \RR stars are
very small, and not significant; they are well within the uncertainties in the
atmospheric parameters. The average difference between Fe abundances obtained
from neutral and singly ionized lines is $0.05\pm 0.03$~dex ($\sigma
=0.10$~dex), in the sense that abundances given by neutral lines are smaller,
that is, this difference has the opposite sign of that obtained for the Sun.
However, if we remove from this comparison ST~Boo and VX~Her, for which no B-W
analysis is available, the mean difference drops to $0.01\pm 0.02$~dex ($\sigma
=0.06$~dex, 8 stars). If differential abundances with respect to the Sun are
considered, [Fe/H] values derived from neutral lines are on average lower than
those derived from singly ionized lines by $0.11\pm 0.03$~dex. The mean slope
in the abundance {\it vs} excitation potential is $0.000\pm 0.008$~dex/eV
($\sigma =0.024$~dex/eV), and it is consistent with the value obtained for the
Sun. The trend with ionization could be eliminated by raising all \teff\ for
the program stars by 70~K, which is well within the uncertainties of the
adopted temperature scale. The observed scatter is compatible with internal
errors of $\pm 80$~K in \teff\ (due e.g. to errors of 0.015~mag in $E(B-V)$)
and to errors of $\pm 0.15$~dex in $\log g$, once the error in the
determination of abundances from individual lines (0.15~dex) is taken into
account. Internal errors in the abundances deduced from neutral and singly
ionized Fe lines are 0.08 and 0.09~dex respectively: then, it seems that the
best estimate of [Fe/H] (to be compared e.g. with photometric data) is obtained
by averaging these two values.

Table~11 reports the effects on the abundance derived for each element of the
variation of the atmospheric parameters within their quoted uncertainties, as
well as the effects due to a variation of the solar reference abundances and to
the use of a turbulence velocity varying with optical depth. This table thus
provides an estimate of the systematic errors affecting our abundance analysis.
These systematic errors add to 0.12 and 0.08~dex for FeI and FeII respectively,
on average are of about 0.11~dex, and only in a few cases (O~I and Al~I) can be
as large as 0.26~dex.

\subsubsection{ Comparison with results from statistical equilibrium
computations}

Combining possible systematic errors in the temperature and gravity scales and
in the $gf$s, we get an upper limit of $\sim 0.2$~dex for the systematic
difference between abundances given by neutral and singly ionized lines, due to
some overionization; on the other hand, there is no evidence for departures
from Boltzmann excitation equilibrium at $\sim 0.1-0.2$~dex level. These
observational data can be compared with predictions from statistical
equilibrium
calculations. The aim here is to constrain the relevance of collisions in this
kind of computation. Once properly calibrated, these cross sections can thus
be used in computations appropriate for stars in other evolutionary phases
(subdwarfs, metal-poor RGB stars). To this purpose we used the 61-level Fe~I
model atom by Gratton et al (1995). This model atom reproduces quite closely
Steenbock's (1985). Given the uncertainty in the collisional cross sections,
we adopted a parametric approach: we assumed that the overall transition rate
due to collisions $c$\ is represented by the sum of two terms due to collisions
with electrons ($c_e$) and with neutral hydrogen atoms ($c_H$), respectively.
The last one was obtained following the approach by Drawin (1968, 1969); we
then may write (see Gratton et al 1995):
$$ c_H = k\, f\left( \frac{m_A}{m_H},T, \frac{p_g}{p_e} \right) \,c_e  ,$$
where $m_A$\ and $m_H$\ are the atomic weights for the considered atom (Fe)
and that of hydrogen, $T$\ is the temperature, $p_g$\ and $p_e$\ are the
gas and electron pressures, and $k$\ is a free parameter that we empirically
calibrate in order to model observations for \RR stars. This formalism is
adopted in order to reflect the different degree of uncertainty present in
the cross sections for collisions with electrons (sometimes rather well
known, as in the case of Na, though not of Fe) and with neutral hydrogen atoms
(always very poorly known). Note however that in several cases (including Fe)
$k$\ is better interpreted as an estimate of the overall relevance of
collisions; in fact, we may write:
$$ c = c_e\, \left[ 1+k\,f \left( \frac{m_A}{m_H},T, \frac{p_g}{p_e} \right)
 \right]  .$$
Results of our statistical equilibrium computations made using MULTI code by
Scharmer \& Carlsson (1985) are shown in Figures~7a,b, which compare departure
coefficients obtained with $k=31.6$\ and $k=0$\ for a typical atmosphere of an
\RR star close to minimum (\teff=6200~K, log~g=2.75, [A/H]=$-$1.3). A large Fe
overionization is present deep in the atmosphere in the second case, while the
overionization is significant only at shallow optical depths (above the region
of formation of most lines) in the first case. We then synthesized line
profiles using populations calculated by this statistical equilibrium (non-LTE)
computations, and then reanalyzed the resulting equivalent widths by assuming
LTE. Figure~8 shows the differences between abundances obtained from LTE
analysis of neutral and singly ionized Fe lines as resulting from this
exercise, for different values of $k$. The range allowed for these differences
by observational errors and uncertainties in the adopted atmospheric parameters
is shown by the shaded region in Figure 8. From this figure, we derive $k\geq
30$. However, given the large uncertainties in all collisional cross sections
for Fe, the correct interpretation of this result is that collisional cross
sections are much larger than originally assumed in our Fe I model atom, rather
than that collisions with neutral H atoms are much more important than
collisions with electrons.
Anyway, if the
collisional cross sections determined by this procedure are used in the
computation of statistical equilibrium for subdwarfs and metal-poor RGB stars,
it turns out that corrections to LTE abundances for Fe are negligible for
these stars, supporting results of classical LTE analyses. The results of
detailed computations will be given in Gratton et al (1995).

\subsubsection{ Comparison with previous results }

We have compared our abundances with those derived previously for the same
stars.
We have four stars in common with Butler (1975) and Butler \& Deming (1979).
Our abundances for these variables are systematically lower by a
constant offset; the differences are $0.18\pm 0.04$~dex ($\sigma=0.07$~dex),
and $0.17\pm 0.07$~dex ($\sigma=0.13$~dex), with Butler and Butler \& Deming
respectively. Butler \& Deming abundances were derived using
only singly ionized lines, thus a more appropriate comparison is made
with our abundances from Fe~II lines. In this case the mean difference is
$0.14\pm 0.08$~dex ($\sigma =0.16$~dex). A direct comparison of the atmospheric
parameters is not possible since Butler (1975) used a curve-of-growth method,
while Butler \& Deming spectra were taken at a very different phase due to
observational limitations. We have estimated temperatures at phases
corresponding to Butler \& Deming spectra using our color-temperature
calibrations. We found that these temperatures are about 500~K cooler than
Butler \& Deming's for SW~And and X~Ari, while the \teff\ of RR~Cet is 30~K
cooler. The adoption of such cooler temperatures would result in Fe abundances
$\sim$0.2~dex lower than those obtained by Butler \& Deming,
bringing them into agreement with the present paper Fe abundance estimates.

We may also compare our Fe abundances for VY~Ser with those determined by
Carney \& Jones. As mentioned in Section~2.3, $EW$s and atmospheric parameters
adopted by these authors are close to ours. The agreement is good: Carney \&
Jones found [Fe/H]=$-$1.82 and $-$1.73 from neutral and singly ionized lines;
our values are [Fe/H]=$-$1.78 and $-$1.64.

\subsubsection { Abundances at maximum light }

For one of the program variables (SW And) we have acquired spectra at phases
close to both minimum and maximum light. The latter were taken to study
the H-emission which is often associated with the hump in the light
curve of the ab-type RR Lyraes, a feature that roughly
corresponds to $\phi \sim$0.9 of the pulsation.
\par
To address a referee's comment we
analyzed and compared the abundances derived from the two sets of spectra.
However, it should be mentioned that
serious concern arises when the region close to maximum
light is used for abundance analysis of \RR stars. This is, in
fact, a region of fast acceleration where magnitude, colors, radial velocity
and hence temperature and gravity of the variable, change by a large amount in
a very short time. Shock waves propagate through the continuum and line
forming regions during this part of the pulsation cycle, one of which, the so
called main-shock (Fokin 1992), is associated with the hump in the light curve.
The energy released by the main shock produces H-emission and UV-excess that
can
still be detected some time after the hump itself (Preston \& Paczynsky 1964,
Gillet \& Crowe 1988, Clementini et al 1994b). Under such circumstances strong
departure from LTE conditions and moreover important modifications of the
atmospheric structure, are expected to take place, making the use of LTE
model atmospheres in hydrostatic equilibrium probably
no longer appropriate for high precision abundance analysis.
\par
No H-emission has been detected, however, in our spectra of SW And close to
maximum light. Indeed  Preston \& Paczynsky (1964) found for SW And the weakest
and shortest-lived H$\gamma$-emission, among the five ab-type RR Lyraes
they studied.

Effective parameters at phases close
to maximum were derived following the same precepts adopted for spectra at
minimum light. However, we must point out that in the phase range 0.85-1.00
\teff and log~g changed by large amounts (up $\sim 500$~K, $\sim 0.5$~dex)
while each single exposure was acquired, since typical exposures were of about
20-25 minutes, in order to get a reasonable S/N ratio: therefore
the adoption of a
single value of \teff and log~g during this part of the light-curve might be
questioned. The adopted values are those corresponding to the phase of median
energy within the range covered by each exposure. \teff's at these phases are
somewhat different from the values at mean exposure, depending on the part of
the light-curve sampled; however, this choice reduced significantly the scatter
in abundances deduced from individual spectra. The first part of
Table~13 lists Fe abundances
obtained from individual spectra corresponding to the region around maximum
light : for comparisons, the values obtained at
minimum light are also given. The average Fe abundances obtained from spectra
at maximum (log~n(Fe)=7.50 and 7.46 for Fe I and II respectively) are very
close to the values obtained at minimum light (7.45 and 7.50); the r.m.s.
scatters of individual values (0.10 and 0.05~dex) are quite small. However,
inspection of Table~13
reveals that average Fe abundances from individual
spectra are correlated with the slope of abundances deduced from Fe I lines
with excitation potential. This suggests that the adopted \teff's are not the
best values: it should be noticed that for spectra taken during the raising
part of the light-curve, rather large errors in \teff's deduced from colors may
be due to even small errors in the adopted phases, in view of the large time
derivatives of colours. The second part of Table~13
gives the Fe abundances
obtained when \teff's deduced from colors are replaced by those obtained by
forcing the slope of abundances deduced from Fe I lines with excitation
potential to have the same value over the whole light curve. The average Fe
abundances obtained with these \teff's (7.47 and 7.46) are even closer to the
values at minimum light than those obtained with \teff's deduced from colors;
furthermore, the scatter of values obtained from individual spectra is now
reduced to 0.04 for both Fe~I and Fe~II, which is equal to the internal error
determined from the line-to-line scatter in each spectrum. The small trend with
phase in the difference between abundances from neutral and singly ionized
lines might be attributed to small ($\sim 0.2$~dex) errors in the adopted
gravities: this errors are within the uncertainties of numerical integration of
the radial velocity curve during the fast raise at maximum.

The extension of this comparison to other elements shows that abundances
deduced
from spectra taken at maximum light usually agree very well with those obtained
at minimum light: for most elements, differences are $<$0.10~dex if \teff's are
derived from the light curve, and $<$0.08~dex if they are deduced from the
equilibrium of ionization for Fe~I. Larger differences (in the sense abundances
at minimum minus abundances at maximum light) were obtained only for Na
($-$0.24), Ca ($-$0.13), and Mg (+0.21): these values were obtained
using \teff's from Fe I excitation, but they are very similar to those obtained
using \teff's from colors. Since these discrepancies are present only for
neutral species, and correlates with ionization potential, we are inclined to
attribute them to small deviations of the model atmospheres from real ones.
Indeed, we think that the agreement between the abundances deduced from spectra
obtained at the extremes of the light curve is astonishing, once the large
variations of the atmospheres during these exposures are considered. On the
whole, we think these comparisons strongly support the present abundance
analysis.

\subsection{ $\alpha-$elements }

\noindent
Table~14 collects the results for the abundances of the
$\alpha-$elements. The last column gives the average overabundances of all the
elements listed in the table for each star, while the mean overabundances of
$\alpha-$elements in metal-poor \RR stars ([Fe/H]$<-1$) are listed in
Table~15. For comparison, the mean overabundances derived from
accurate abundance analysis of other groups of metal-poor stars are also given
in the last three columns of Table~15 :\\
(i) Column~5: 9 extremely metal-poor main sequence and subgiant stars
recently measured by Nissen et al (1994)\\
(ii) Column~6: 20 metal-poor dwarfs and subgiants observed by Magain (1989) and
Zhao \& Magain (1990)\\
(iii) Column~7: 13 metal-poor dwarfs and giants analyzed by Gratton \& Sneden
(1991) and Carretta et al (1995).

Tables~14 and 15 show that all the $\alpha-$elements are
overabundant in the observed metal-poor \RR variables, the mean
overabundances being similar to those observed in non-variable less evolved
stars (see also Figure~9). The star-to-star scatter of the average
$\alpha-$element overabundances amongst metal-poor stars
is small (0.06~dex) and is compatible with the
hypothesis that all stars share the same enhancement of these elements. The
scatter in element-to-element ratios is not large, and likely not significant.

Finally, it is worth mentioning the good agreement between the abundances
obtained from neutral and singly ionized lines for Si: the mean difference is
$-0.04\pm 0.05$~dex ($\sigma=0.12$~dex, 6 stars). This fact supports the LTE
analysis. Conversely, our LTE analysis yields slightly larger Ti
abundances from Ti~II lines than those obtained from Ti~I lines (the mean
difference is $0.14\pm 0.06$~dex, $\sigma=0.18$~dex, 9 stars), supporting a
(small) Ti overionization. A similar suggestion was hinted by the analysis of
metal-poor RGB stars by Gratton \& Sneden (1991).

\subsection{ Fe-group and heavy elements }

\noindent
Tables~16 and 17 list abundances for the Fe-group and
heavy elements ($Z>28$) in individual stars, while average overabundances in
metal-poor ([Fe/H]$<-1$) variables are given in Table~18. For
Sc~II, Mn~I, and Cu~I lines, the hyperfine structure was considered in detail
(see Gratton 1989, Gratton \& Sneden 1991, and Sneden et al 1991 for
references). For comparison, abundances derived for metal-poor main-sequence
and red giants (Gratton 1989, Gratton \& Sneden 1991, 1994, and Sneden et al
1991) are listed in Table~18. Main results are:\\
(i) Sc and Y abundances scale as Fe, in agreement with less evolved cooler
stars. However, we obtained very low abundances for metal-rich variables (see
next subsection).\\
(ii) Cr abundances derived from Cr~II lines also scale as Fe. There is some
indication of overionization of Cr, since neutral lines give systematically
lower abundances (by $0.25\pm 0.06$~dex, $\sigma=0.20$~dex, 10 stars). A
similar result was found by Gratton \& Sneden (1991) amongst the most
metal-poor red giants.\\
(iii) Mn is found to be severely underabundant in metal-poor RR~Lyraes; a less
pronounced Mn underabundance was found in red giants (Gratton 1989). The larger
effect seen in RR~Lyraes may either be real (and in this case it would suggest
a smaller contribution to nucleosynthesis by type~I SNe), or may be the effect
of some systematic deviation due to atmospheric effects or departures from LTE
in the formation of the Mn resonance lines (see also Gratton 1987).\\
(iv) The observed Ba abundances fit well in the pattern observed in other
metal-poor stars.

\subsection{ Sc and Y }

\noindent
Abundances of some of the ionized species (Sc, Y, and rare earths) in
metal-rich
stars appear anomalous when compared to solar ones (which are typical of
metal-rich objects). By combining data for SW~And and V445~Oph, we get the
results listed in Table~19. Inspection of this table shows that the
abundances of Nd~II, Ce~II, Y~II, and Sc~II are
anomalous \footnote{Abundances from Nd II and Ce II lines were derived only for
SW~And.}: no reasonable combination of errors in the atmospheric parameters
may justify these results, which are confirmed by a comparison with synthetic
spectra for some of the features of these species. For the remaining elements,
there is no clear evidence for anomalies. Ca~II is listed in Table~19
for its interest in the derivation of metallicities by means of
the $\Delta $S and
\CaII methods. Although we have not derived Ca abundances from singly
ionized lines, the {\it normal} abundances obtained from neutral lines argue in
favour of a {\it normal} strength for the Ca~II lines. It is difficult to
explain the underabundances of Nd~II, Ce~II, Y~II, and Sc~II by means of a
nucleosynthetic argument; it is then interesting to explore the possibility
that these anomalous abundances are due to an atmospheric effect that falsifies
the results of our LTE analysis. We suspect that this mechanism may be
related with the propagation of the shock waves connected with the pulsation,
because non variable stars (but also Cepheids) with
atmospheric parameters similar to \RR variables do not display Sc
and Y underabundances (Luck \& Bond 1989; Luck et al 1990). $\delta$~Del
stars (which are only slightly warmer than RR~Lyraes at minimum: \teff$\sim
7000$~K {\it vs} \teff$\sim 6300$~K) exhibit even overabundances of heavy
elements. Unfortunately, accurate atomic data (cross sections etc.) for these
elements are scarce, and do not allow to draw a very detailed model. We suggest
that photoionization of these elements throughout most of the atmosphere might
be due to photons emitted in the Lyman lines produced by the shock waves which
propagate in the atmosphere of these variables (see Fokin 1992). We do not
expect overionization of Ti~II, Fe~II, Si~II, and Cr~II, requiring Lyman
continuum photons, which are abundant only in the very small regions
immediately behind the shocks. Since Nd, Ce, Sc, and Y are more that $10^{9}$\
less abundant that H, this process is very minor from the point of view of the
net energy balance: the total energy required to keep these elements twice
ionized in the atmosphere of RR~Lyraes is $\sim 3\,10^{19}$~erg/s for a mass
loss rate of $\sim$ $10^{-11}~\Msol$ /yr (the energy required to ionize twice
all Sc and Y in the atmosphere is $\sim 3\,10^{25}$~erg). A priori, it is not
clear why this mechanism is efficient for Sc, Y, Ce, and Nd, while it does not
work for Ca and Ba, why it is not balanced by recombinations, nor why it
is not present in more metal-poor stars in our sample. Perhaps, this peculiar
behaviour for Ca and Ba may be due to the small value of the photoionization
cross section from the fundamental level of atoms and ions having an alkaline
structure; these species have also a rather large jump to the first excited
levels (3.13~eV for Ca~II, 2.72~eV for Ba~II) ensuring that these last are
scarcely populated. On the other hand, photoionization cross sections are much
larger for atoms and ions having an alkaline-earth structure, like Sc~II and
Y~II (which also have a rather large number of low energy levels). Both
hydrodynamical and statistical equilibrium computations are required to give a
better insight into this issue.

\subsection{ Mixing and abundances of light elements (C, O, Na and Al) }

As mentioned in the Introduction, an anticorrelation between Na and O
overabundances has been found for RGB stars in metal-poor globular clusters
(see e.g. Sneden et al 1994 and references therein). Langer et al (1993)
discussed various possible explanations for this anticorrelation, and
reached the conclusion that the most favoured mechanism
is the dredge-up of NO-cycle processed
material related to meridional circulating currents, possibly activated by core
rotation (Sweigart \& Mengel 1979). No such anti-correlation has been observed
for RGB stars in metal-rich globular clusters (Brown \& Wallerstein 1992,
Carretta \& Gratton 1992, Sneden et al 1994). This metallicity dependence is
commonly attributed to the stronger molecular weight barrier existing in
metal-rich giants (Tassoul \& Tassoul 1984), which inhibits the mixing.
Observations of light elements (C, O, Na and Al) in \RR stars are of great
interest because these stars are in an evolutionary phase following the RGB,
and thus must have the
envelope composition inherited from this previous evolutionary stage.
Various features of C, O, Na, and Al are observable in our spectra. We measured
the $EW$s of the O~I permitted triplet in the near IR (7771-7774~\AA) in all
program \RR stars; however, these lines are very strong and their formation is
likely affected by significant departures from LTE. The much weaker permitted
triplet at 6155-6158~\AA\ was observed in the two most metal-rich stars (SW~And
and V445~Oph). Since these lines are rather weak, departures from LTE are
expected to be small: however these lines are blended at the resolution of our
spectra, so that we derived O abundances by comparison with synthetic spectra.
The line list used in these computations includes those lines from Kurucz \&
Peytremann (1975); wavelengths and $gf$s for the stronger lines were adjusted
using updated literature data and a comparison with the solar flux spectrum by
Kurucz et al (1984). For these metal-rich stars, we were also able to measure
$EW$s for lines of the C~I multiplet at 7111-7120~\AA. Finally, Na and Al
abundances were measured from spectral synthesis of the doublet at 5682-88~\AA\
(which unfortunately are too weak to be measured in the spectra of the most
metal-poor \RR stars), of the D lines, and of the Al resonance line at
3961~\AA. These syntheses were obtained following the same rules adopted for
the O lines. Figure~10 shows a comparison between synthetic and computed
spectra for a few stars. Results for C, O, Na and Al abundances are summarized
in Table~20. Note that $EW$s for the O~I IR triplet are more
uncertain than average in our analysis, due to difficulties in eliminating the
strong interference fringes present in the spectra. We suspect that the large O
abundances for VX~Her and VY~Ser are due to this cause. Further, interstellar D
lines are blended with stellar features in the spectra of a few of our
variables, which had a low geocentric velocities when observed. No Na abundance
could be drawn from the D lines when this blending was too severe (V445~Oph and
SW~And), while results are quite uncertain for other stars (V440~Sgr, RR~Lyr,
and X~Ari). Uncertain values are marked by a colon in Table~20.

Departures from LTE were considered through detailed statistical equilibrium
calculations for O and Na lines, using the same procedure adopted for Fe (see
Section~4.1.3). The (new) O and Na atomic models are fully described in Gratton
et al (1995). As for Fe, we calibrated empirically the cross sections for
collisions with H~I atoms; results from these calculations are shown in
Figures~11a,b. Here, we forced agreement between the abundances derived from
weak and strong lines (the multiplets at 7771-74 and 6155-58~\AA\ for O; the
doublets at 5893 and 5685~\AA\ for Na) in the non-LTE analysis for those stars
for which all these features were measured.  As in Figure 8 shaded regions in
Figures 11a,b correspond to the range allowed by observational errors and
uncertainties in the adopted atmospheric parameters.

For Oxygen, we ran several test cases for the Sun; our results agree fairly
well with those by Kiselman (1991, 1993). As anticipated, we found that the IR
permitted triplet is much stronger than expected in LTE,
while departures from
LTE are negligible for the much weaker lines at 6155-58~\AA\ in the case of RR
Lyraes. Our statistical equilibrium computations also show that the Na~D lines
are much stronger than predicted by LTE, while, as expected, the subordinate
doublet at 5682-88~\AA\ forms close to LTE. This is due to the combination of
the photon suction effect described by Bruls et al (1992), and of the depth of
formation of the lines (the D~lines are heavily saturated in the spectra of RR
Lyrae). A more complete discussion will be given in Gratton et al (1995).
We then estimated corrections to LTE abundances as a function of
line strength, for model atmospheres representative of variables having
different metallicities (see Figures~12a,b). The abundances listed in
Table~20 include these corrections.

The results of Table~20 agree fairly well with the predictions by
canonical mixing (first dredge-up: Iben 1965); all metal-poor variables have
large O overabundances, while Na and Al are overdeficient with respect to Fe.
We regard as doubtful the very large O overabundances we
found for VX~Her and VY~Ser.
These results reproduce the typical pattern observed in less evolved metal-poor
stars (see Figure~13). Our Na abundances are even slightly smaller than
those observed by others, who however did not include any non-LTE correction.
Therefore, we do not find any evidence for an anticorrelation between Na and O
overabundances in our sample. This results is not totally unexpected; in fact
no
{\it bona fide} O-poor, Na-rich star has been found among field first ascent
giants (see Sneden et al 1992). Their absence suggests that the environment
(possibly in some indirect way) plays a main r\^ole in the mechanism causing
the anomalous abundances observed in cluster stars (Langer et al 1993).

\section{ DISCUSSION }

\subsection{A new calibration of the \DS index}

Our new iron abundances have been used to re-calibrate the \DS
metallicity index.
The newest and most complete \DS compilations for field \RR stars are
those of B92 and SKK94. The two systems are virtually identical; in fact if we
compare \DS values for the stars in common between the two samples (47 stars)
we find that, after omitting the very discrepant UU~Cet, the linear correlation
between the two systems is:
$$ \Delta S_{\rm B92} = 0.97 (\pm 0.03) \Delta S_{\rm SKK94} + 0.09 (\pm 0.11)
$$
or\\
$$ \Delta S_{\rm SKK94} = 0.98 (\pm 0.03) \Delta S_{\rm B92} + 0.21 (\pm 0.11)
$$
To derive the \DS-\feh calibration we took  therefore the mean values between
$\Delta S_{\rm B92}$ and $ \Delta S_{\rm SKK94}$. Using for the \feh values the
average of our Fe~I and Fe~II abundances we obtained the new calibration:
$$ [{\rm Fe/H}] = -0.217 (\pm 0.015) \Delta S\ + 0.04 (\pm 0.05) ~~~(5)$$
which is the mean of the two least-squares solutions made by exchanging the
dependent and the independent variables. The correlation coefficient is r=0.98
with $\sigma$=0.16 (10 stars): this $\sigma$ is the accuracy of the
[Fe/H] value derived for a single star using \DS and our calibration. The new
\DS {\it vs}
\feh calibration is shown in Figure~14a (solid line).
Errors are 0.08~dex in [Fe/H] and
0.7~units in \DS (assuming for both \DS systems an uncertainty of one unit in
\DS).

Our new calibration is steeper and has a larger zero-point than Butler's
(1975; dotted line in Figure 14a). This results
in a different metallicity scale in which old [Fe/H]
determinations are now {\it stretched}, so that we obtain larger abundances for
metal-rich stars and lower abundances for metal-poor ones, the differences
being almost negligible in the intermediate regime of metallicity ($\Delta S
\simeq$ 4.7, [Fe/H] $\sim -1.0$). For $\Delta S =-1$\ and $\Delta S =12$\
we have that $\Delta$[Fe/H]=[Fe/H]$_{\rm our}-$[Fe/H]$_{\rm Butl.}$=+0.33 and
$-$0.41, respectively.

A more accurate calibration of the \DS\ index can be obtained by combining our
[Fe/H] values and those of Butler and coworkers (average of Butler 1975 and
Butler \& Deming 1979), corrected for the systematic difference with respect to
our [Fe/H] values ($-0.18$\ and $-0.17$~dex respectively, see Section~4.1.4).
Although the corrected Butler's [Fe/H] values are less accurate than ours, we
nevertheless obtain a better distribution, particularly in the range
$0<\Delta S<5$.
We get a very tight relationship between \DS\ and [Fe/H] (see Figure~14b). The
resulting calibration is:
$${\rm [Fe/H]} = -0.204(\pm 0.012) \Delta S - 0.10(\pm 0.19), ~~~~(6) $$
which is the average relation obtained by exchanging dependent and independent
variables. It should be noted that the slope of this relation agrees within the
errors with that of the original Butler calibration ($-0.16\pm 0.02$, m.e.
rather than standard deviation). The improvement is due to both better [Fe/H]
values and better $\Delta S$. A steeper relation (slope of $-0.18\pm 0.05$) was
obtained already by B92, still using Butler's [Fe/H] but new \DS. Of course,
eq. (6) agrees well with eq. (5).
\subsubsection{ Comparison with the Globular Cluster metallicity scale }

The new calibration of $\Delta S$\ may be compared with the results
obtained by using RR~Lyraes in globular clusters, which have [Fe/H] determined
from high dispersion spectroscopic observations of RGB stars. We collected
[Fe/H] values from Gratton et al (1989), Leep et al (1987), Brown et al
(1991, 1992), Sneden et al (1991, 1992, 1994), Kraft et al (1992, 1993),
Peterson et al (1990), and Fran\c cois (1991). An inter-comparison between
these different sets of data shows that the [Fe/H] values of Gratton et al
(which refer to the largest number of clusters) are in good agreement with
those of Leep et al, Brown et al, and Fran\c cois. On the other hand,
[Fe/H] values obtained by Sneden and coworkers using large samples of stars
in each cluster, observed at very high resolution and S/N, are systematically
larger by $\sim 0.2$~dex than those obtained by
the other authors. The reason for this systematic difference
is not entirely clear; we suspect it is due to a different reduction to the
solar Fe abundance. However, it is not clear what procedure should be
preferable, in view of the difficulties related to the way collisional damping
(which is important when solar abundances are used) is handled.
Hereinafter, we simply averaged abundances obtained by different authors for
individual clusters. We verified that none of our conclusions is modified if
systematic corrections as large as 0.2~dex are applied to the various sets of
[Fe/H] for globular clusters.

We then calibrated the $\Delta S$\ index using only data for globular
clusters. To this purpose, we used the $\Delta S$\ values collected by
Costar \& Smith (1988) and references therein. The regression line between
[Fe/H] and $\Delta S$, (where the usual average between values obtained
exchanging dependent and independent variables was applied), is:
$${\rm [Fe/H]} = -0.168(\pm 0.016) \Delta S - 0.22(\pm 0.13) ~~~~(7)$$
The difference between the slopes of eq. (6) and (7) is significant
at about the 2~$\sigma$ level. Although not highly significant,
this difference does call for an explanation.

A direct comparison between results for field and globular stars is given in
Figure~14c, where [Fe/H] values are plotted against $\Delta S$\ both for field
and globular cluster RR~Lyraes. While there are no RR~Lyraes in metal-rich
globular clusters, the two different sets of data cannot otherwise be
distinguished.
This suggests that the calibrations obtained from field and cluster stars
differ either simply due to random variations, or because the relation between
$\Delta S$\ and [Fe/H] is not linear, and the metallicity distributions of the
two groups of objects are not identical. Present data are not good enough to
settle this issue: in fact, while data plotted in Figure~14c may suggest a
smaller slope in the range $4<\Delta S<10$, we found that the slope obtained
using all objects (field and globular cluster stars) in this range ($-0.159\pm
0.038$) is not statistically different from the value obtained using all
objects regardless of $\Delta S$. The average calibration we obtain
considering both field and cluster variables is:
$${\rm [Fe/H]} = -0.194(\pm 0.011) \Delta S - 0.08(\pm 0.18) ~~~~(8)$$
which can be considered our final, adopted calibration.  More, accurate
abundance determinations would be welcomed.

We finally remark that a non-linear relation between $\Delta S$\ and [Fe/H]
could be predicted theoretically. In fact, Manduca (1981) showed that the
$\Delta S$\ index is expected to saturate at low metallicities ([Fe/H]$\sim
-2.5$), while the [Ca/Fe] ratio is known to be lower in metal-rich stars
than in metal-poor, population II stars (as discussed in Section~4.2).
However, if the calibration of $\Delta S$\ {\it vs} [Fe/H] is indeed affected
by
the trend of the [Ca/Fe] ratio with [Fe/H], considerable care should be
devoted in extending the calibration inferred from stars in the solar
neighborhood to RR~Lyraes observed in environments where the run of the
$\alpha-$elements with respect to
Fe is expected to be different, like the galactic bulge
or the Magellanic Clouds.

\subsection{A new calibration of the W$'$(K) index}

The other technique to derive the metallicity from low-resolution spectra of RR
Lyrae stars is the equivalent width of the Ca II K line corrected for
interstellar contribution (Clementini et al 1991). Using the present new
metallicities, and $W'(K)$ values and relative errors  from Clementini et al
(1991), (we have 8 objects in common with that paper, but have omitted VY~Ser,
whose $W'(K)$ value is rather uncertain), a least-squares fit weighted both in
$W'(K)$ and \feh gives:
$$ {\rm [Fe/H]} = 0.65(\pm 0.17)W'(K) - 3.49(\pm0.39) ~~~(9)~~
{(\rm 7~objects)}$$
\noindent
where, according to the procedure by Clementini et al (1991), [Fe/H] values
were derived from Fe II abundances. The new $W'(K)$ {\it vs} \feh calibration
is shown in Figure~15 (solid line).
Error bars of the \feh values correspond to 0.09~dex
(see Section~4.1.2).

Same as for \DS, the new $W'(K)$ {\it vs} \feh calibration is steeper and has a
larger zero-point than Clementini's et al (1991; dotted
line in Figure 15). This translates in a larger
metal abundance derived for metal-rich RR~Lyraes and lower abundances for
metal-poor ones. For $W'(K)$=1 and $W'(K)$=6.5 we have that
$\Delta$[Fe/H]=[Fe/H]$_{\rm our}-$[Fe/H]$_{\rm Clem.}$=$-$0.30 and
+0.31, respectively, while we get almost the same metallicity [Fe/H]$\sim
-1.2$, for $W'(K)$=3.5.

\subsection{The M$_V$ -- [Fe/H] dependence}

We have used our new metallicity scale to revise the metallicity dependence of
the absolute magnitude of \RR stars, M$_V$. This is a very crucial step in
deriving the galactic distance ladder and the ages of Galactic Globular
Clusters (GGCs). If accurately known, the slope of this relationship could
allow us to calibrate the relative ages of GGCs, discriminating  between the
$slow$\ and $rapid$\ halo collapse scenarios, while the zero-point would fix
the age of some reference cluster, enabling absolute age determinations.

Our aim is to see if the new metallicity scale has some effect on the slope of
M$_V -$ [Fe/H] dependence. The most up-dated list of B-W M$_V$\ estimates
derived from infrared colors was published by Fernley (1994), who also provided
a re-evaluation of the M$_V$\ values using a perhaps more appropriate value of
the conversion factor between observed and true pulsation velocity $p=1.38$\
than previously adopted (see Fernley 1994 for a detailed discussion of
this issue). Fernley's (1994) sample contains 29 field RR~Lyraes. To these, we
added 8 globular cluster variables on which BW analysis has been performed by
Liu \& Janes (1990b; V2, V15, V32, V33 in M4), and by Storm et al (1994; V8 and
V28 in M5, and V1 and V3 in M92). The relevant data for these stars are
collected in Table~21 where we give the star name (Column~1); \DS
values obtained as the average between B92 and SKK94 values (Column~2); \feh
values and related uncertainties (Column~3 and 4 respectively), from high
resolution abundance analysis or derived from the \DS values in Column~2 and
our \DS-\feh calibration (eq. 8); the absolute magnitudes M$_V$ from
Fernley (1994) for the field stars, listing both the original values (Column~5)
and the values corrected to {\it p}=1.38 (Column~6), and the associated errors
(Column~7: from Skillen et al 1993).

For the cluster variables, M$_V$ values in Column~5 and uncertainties in
Column~7 are from Liu \& Janes (1990b), and Storm et al (1994), respectively.
These M$_V$s were derived using {\it p} =1.31 and 1.30. According
to Table~1 of Fernley (1994) the M$_V$ values of the M4 variables were made
brighter by 0.11 mag and those of M5 and M92 were made brighter
by 0.13 mag to transform them to $p=1.38$\ (see values in Column~6).
Metallicity for the M4 variables is the average value derived in Paper~I. For
M5 and M92 we used [Fe/H]=$-$1.17 and [Fe/H]=$-$2.25, as
derived by Sneden et al (1994, 1992) from high resolution spectroscopy of RGB
stars. The adoption of a metallicity slightly different from that used to
derive the M$_V$ values via B-W analysis should not affect our conclusions
since all the quoted B-W M$_V$ estimates were obtained by using the $V-K$ color
that is almost metallicity independent.

The data listed in Table~21 are plotted in Figure~16, with the original M$_V$
values in panel (a) and the value corrected to {\it p}=1.38 in panel (b).
Different symbols in Figure~16 refer to: field RR~Lyraes (open and filled
squares, the last ones being the objects studied in the present paper, for
them, we have used our \feh estimates); RR~Lyraes in M5 and M92 (open triangles
and open circles, respectively) according to Storm et al (1994); and RR~Lyraes
in M4 (filled triangles) according to Liu \& Janes (1991). Asterisks mark
field variables that are claimed to be evolved by Jones et al (1992), Cacciari
et al (1992), and Skillen et al (1993). We note that, a part from SS Leo and
the two variables in M92, all the other stars do not seem to deviate
significantly
from the general distribution in panels (a) and (b). Moreover BB Pup is
underluminous, while evolution off the ZAHB should make the star overluminous.

Omitting SS Leo and the M92 stars, (see the discussion on the reasons for
omitting evolved stars when calculating the M$_V$ {\it vs} \feh relation in
Cacciari et al 1992), a least-squares fit weighted both in M$_V$ and \feh
gives:
$${\rm M}_V=0.20(\pm 0.03){\rm \feh}+1.06(\pm 0.04)~~~(r=0.72~~~\rm
34~objects)~~~~
(10a)$$
which was obtained using the M$_V$ values in Column~5 of Table~21,
and:
$${\rm M}_V=0.19(\pm 0.03){\rm \feh}+0.96(\pm 0.04)~~~(r=0.70~~~\rm
34~objects)~~~~
(10b)$$
which was obtained using the corrected M$_V$ values in Column~6 of
Table~21. These best-fit regression lines are plotted on the data in
Figure~16.

Analogous best-fit calculations, but omitting all the supposed evolved stars,
(SU Dra, W Tuc, SS Leo, BB Pup and the M92 variables) give:
$${\rm M}_V=0.17(\pm 0.03){\rm \feh}+1.03(\pm 0.04)~~~(r=0.76~~~\rm
31~objects)~~~~
(10c)$$
and
$${\rm M}_V=0.17(\pm 0.03){\rm \feh}+0.93(\pm 0.04)~~~(r=0.73~~~\rm
31~objects)~~~~
(10d),$$
respectively. The adoption of the new metallicity scale does not yield
significant changes in the slope and zero-point of the M$_V$ {\it vs} \feh
relation. Indeed, within the quoted uncertainties, slopes and zero-points of
eq. 10a and 10c agree very well with those found by Liu \& Janes
(1990a,b), Jones et al (1992), Cacciari et al (1992), and Skillen et al (1993);
therefore the conclusions drawn in those papers, in particular about the ages
of the Galactic globular clusters, apply here as well. According to Fernley
(1994), a zero-point brighter by 0.10 mag is found when using M$_V$ values
derived for {\it p} =1.38 (see eq. 10b and 10d). However this zero-point
is still $\sim$ 0.08 mag fainter than the zero-points of the M$_V$ {\it vs}
\feh relations predicted by Stellar Evolution and Stellar Pulsation theories
(Fernley 1994).

On the other hand, the distributions in Fig.s 16a,b, do not seem to be well
fitted by a single-sloped relationship. Once evolved stars are eliminated, it
appears that there might be two distinct regimes: one, at high metallicity,
requiring a higher slope, and a second one at low metallicity, which would
better be fitted with a much lower slope. If true, this result would not be
entirely unexpected, since Castellani et al (1991) have already pointed out
(from a theoretical point of view) that there is no reason, in principle, to
believe in a linear dependence of the HB luminosities on metallicity. Their
point is that at lower metallicities this dependence is much smaller than in
the higher metallicities regime, where the theoretical slope rises
progressively, approaching the Sandage's (1993a,b) pulsational value. Our
analysis seems to give some support to this prediction; however, a much larger
data sample (particularly B-W M$_V$ determinations) are requested to definitely
settle this question.

\section{ CONCLUSIONS }

\noindent
We derived abundances for 21 species from moderately high-resolution, high
S/N, visible spectra of 10 field {\it ab}-type \RR stars. The main purposes
of this study were:
\begin{itemize}
\item to provide new, updated calibrations of the \DS (Preston 1959) and \CaII
  (Clementini et al 1991) indices to be used in galactic and extragalactic
  studies.
\item to discuss the composition of HB stars, comparing it to that of less
  evolved low-mass stars.
\item to compare the observed line strengths with predictions from statistical
  equilibrium calculations, in order to predict the relevance of departures
  from LTE in the spectra of subdwarfs and RGB stars, where non-LTE effects
  are expected to be smaller than for RR~Lyraes.
\end{itemize}

In order to achieve these results, we put particular emphasis on a precise
estimate of the atmospheric parameters independently of excitation and
ionization equilibria. To this purpose, we selected stars having accurate
photometric and radial velocity data, and made observations close to minimum
phase, when the atmospheres are quite stable and shock waves are out of the
line forming regions. Photometric reddening estimates for the program stars
were carefully examined, and compared with other determinations. Dereddened
colors were then used to infer effective temperatures at the phase of
observations using a new temperature scale determined from literature Infrared
Flux Method measures of subdwarfs, and the Kurucz (1992) model atmospheres,
following a procedure similar to that recently adopted by King (1993).
Gravities were derived from the radial velocity curve of the variables. The
applicability of Kurucz (1992) model atmospheres in the analysis of RR~Lyraes
at minimum light was briefly analyzed: we found that they are able to reproduce
colors, excitation and ionization equilibria as well as the wings of
H$_{\alpha}$. The solar comparison abundances were carefully determined. In
particular, a new solar Fe abundance was obtained from a group of weak Fe~I
lines having accurate $gf$s (Bard \& Kock 1994). We obtained a value of $\log
\epsilon(Fe)=7.52$, in good agreement with the value obtained from meteorites
and Fe~I lines.

The main results of our abundance analysis for RR~Lyraes are:
\begin{itemize}
\item The metal abundances of the program stars span the range
$-2.50<$[Fe/H]$<+0.17$.
\item Lines of most elements are found to form in LTE conditions; there are a
few significant exceptions, that are mentioned below. In particular, we found
that the equilibria of excitation and ionization are very well achieved for Fe
lines. This result is compared with predictions from statistical equilibrium
calculations, using a rather extended Fe~I model atom, including 60 levels plus
the continuum. We found that the originally adopted collisional cross sections
must be raised in order to reproduce observations. If statistical equilibrium
computations with this rather large collisional cross sections are then
repeated for subdwarfs and metal-poor giants, departures from LTE
are found to be negligible, thus validating the LTE analyses for these stars.
\item On the whole, the composition of RR~Lyraes is similar to that of less
evolved stars of similar \feh: $\alpha$-elements are overabundant by
$\sim$0.4~dex and Mn is underabundant by $\sim$0.6~dex in stars with \feh$<
-1$. Solar scaled abundances are found for most of the other species, except
for the low Ba abundance in the extremely metal-poor star X~Ari
(\feh$\sim-2.5$).
\item Significant departures from LTE are found for a few species: Nd~II,
Ce~II, Y~II and Sc~II are severely underabundant ($\sim$0.5~dex) in metal-rich
variables; Ti~I and Cr~I are slightly ($\sim 0.1 - 0.2$~dex) underabundant
in metal-poor stars. These effects are attributed to overionization. We suggest
that the photoionization of the alkaline earth-like ions is due to emission in
the Lyman lines produced by shock waves that propagate in the atmosphere of
these variables (Fokin 1992). However, it is not clear why this effect is not
seen in more metal-poor stars, and it is not observed for Ca and Ba.
\item Departures from LTE were considered in detail in the derivation of
abundances for the light elements (O and Na). A comparison with weaker lines
(forming close to LTE) showed that significant corrections were required for
the O~I IR triplet and the Na~D lines. We were able to model these non-LTE
effects by means of statistical equilibrium computations. The abundances
obtained from
the non-LTE analysis reproduce very well those observed in less evolved field
stars, suggesting that the mixing during the RGB phase did not involve
ON-processed material for all program stars. Hence, we did not find any
evidence for an O-Na anti-correlation among these field HB-stars, suggesting
that the environment is likely to be somehow responsible for the
anti-correlation found in metal-poor globular cluster stars (Sneden et al
1992).
\end{itemize}

A new calibration of the \DS index for $ab-$type RR~Lyraes was then obtained
using our own \feh, as well as those from Butler and coworkers (corrected to
our system), and from high resolution spectroscopy of globular clusters giants:
$$ {\rm [Fe/H]}= -0.194(\pm 0.011)\Delta S - 0.08 (\pm 0.18) $$
Our new metallicity scale is stretched on both low and high metallicity ends
with respect to Butler's (1975). There is some hint that \DS-[Fe/H]
relation may be not linear: in fact the slope obtained using only stars with
$4<$\DS$<10$\ is slightly smaller than that obtained using all stars. While
this difference is only barely significant, it would not be in contrast with
theoretical expectations. In fact, a saturation of the \DS index at low
metallicities has been suggested by Manduca (1981), while at high metallicities
we expect a larger slope due to the variation of the [Ca/Fe] ratio with [Fe/H]
in
the range $-1<$[Fe/H]$<0$. If this non linearity is confirmed by more
observations, caution should be exerted in the derivation of [Fe/H] from \DS
for
stars having [Ca/Fe] ratios different from that observed in stars of similar
[Fe/H] in the solar neighborhood.

In a similar way, we used our new \feh values to update the metallicity
calibration of the \CaII index. Using the present new metallicities, and
$W'(K)$ values and relative errors  from Clementini et al (1991), a
least-squares fit weighted both in $W'(K)$ and \feh gives:
$$ {\rm [Fe/H]} = 0.65(\pm 0.17)W'(K) - 3.49(\pm0.39)$$

Finally, our new metallicity scale was
used to revise the metallicity dependence
of the absolute magnitude of \RR stars, M$_V$. We found:
$${\rm M}_V = 0.20(\pm 0.03){\rm \feh} + 1.06(\pm 0.04)$$
and:
$${\rm M}_V = 0.19(\pm 0.03){\rm \feh} + 0.96(\pm 0.04)$$
the last being obtained by using M$_V$ values derived for a value of the
conversion factor between observed and true pulsation velocity $p=1.38$\
(Fernley 1994). As expected, the adoption of the new metallicity scale does not
yield significant changes in the slope and zero-point of the M$_V$ {\it vs}
\feh relation. A close inspection of the available data does not exclude the
possibility that there might be two distinct regimes in the M$_V$ {\it vs} \feh
plane: one, at high metallicity, requiring a higher slope, and a second one at
low metallicity, which would better be fitted with a much lower slope. A larger
data sample (particularly B-W M$_V$ determinations) would be requested,
however, to definitely settle this question.

\section*{Acknowledgments}

We wish to thank Dr. R.L. Kurucz for kindly making available to us on tape his
new model atmospheres; Dr. M. Carlsson for having provided a version of the
MULTI code for statistical equilibrium computations; Dr. C. Cacciari for her
comments on the color-\teff\ calibration; Prof. B. Gustafsson for many comments
on statistical equilibrium computations; and Dr. A. Bragaglia for help in the
reduction procedures and preparation of figures. We also thank the referee, Dr.
R. Peterson,  for making constructive suggestions concerning the original
manuscript. G.C. and R.G. wish to dedicate this paper to their children Ilaria
and Max.

\section*{Appendix}

In order to derive temperatures from observed colors, we have collected from
the literature the most recent photometric data available for our program
stars. In the following we list the photometric data used for each star, giving
also the equations used to transform original photometries to the
Johnson-Cousins
($BV$\Rc\Ic\Kj) photometric system.

-RR~Cet \\
$BVRI$\ on the Johnson-Cousins system ($BV$\Rc\Ic), and $K$\ photometry on the
CIT system (\Kcit) transformed to Johnson according to eq. (a) (see below),
published by Liu \& Janes (1989).

-RR~Lyr \\
$BV$\ from Fitch et al (1966), $V$\ from Siegel (1982), and $BVRI$\ from
Manduca et al (1981). Manduca et al (1981) photometry is on the Johnson system
defined by Barnes et al (1978) (\Rbm,\Ibm), and was transformed to
Cousins using eq. (b) (see below).

-VY~Ser \\
V photometry from Carney \& Latham (1984), K$_{CIT}$ photometry from Jones et
al
(1988), R$_{BM}$ and I$_{BM}$ from Fernley et al (1990), transformed to
Johnson-Cousins using eqs. (a) and (b).

-X~Ari \\
$V$\Rc\Kcit from Jones et al (1987), $BV$\ from Burchi et al (1993), and
$BV$\Rc\Ic\Kj from Fernley et al (1989) as derived from their Figures~1 and 2
which include also the $VR$\ photometry originally taken from Manduca et al
(1981).

-ST~Boo \\
$BVRI$\ from Clementini et al (1995). $R$\ and $I$\ data are on the
Johnson system
defined by Nechel \& Chini (1980) (\Rnc,\Inc); they were transformed to Cousins
using eq. (c).

-UU~Cet \\
$BV$\Rc\Ic photometry from Clementini et al (1990), and $K$\ photometry on the
ESO photometric system ($K_{ESO}$) from Cacciari et al (1992). $K_{ESO}$ has
been transformed to $K_{J}$  using eq. (d).

-V445~Oph \\
$V$\Kj from Fernley et al (1990), and $BV$\Rbm\Ibm from Barnes et al (1988),
transformed according to eq. (b).

-V440~Sgr \\
$BV$\Rc\Ic from Cacciari et al (1987).

-VX~Her \\
$BV$\ photometry from Stepien (1972), Fitch et al (1966), Sturch (1966) and
$BV$\Rnc\Inc from Clementini (unpublished), transformed according to eq. (c).

\noindent
We have used the following equations to transform original photometries to
the Johnson-Cousins system:\\

\noindent
$(V-K)_{J}=(V-K)_{CIT}-$0.011~~~~(Jones et al 1987)~~~(a)\\
$(V-R)_C = 0.716 (V-R)_{BM}-$0.032~~~$(R-I)_C=0.873\,(R-I)_{BM}$+0.42~~~~
Bessel (1983)~~~(b)\\
$(V-R)_C = 0.757 (V-R)_{NC}-$0.040~~~$(V-I)_C=0.782\,(V-I)_{NC}$~~~~
Bessel (1983)~~~(c)\\
$(V-K)_J = (V-K)_{ESO}-0.013-0.024\,(J-K)_{ESO}$~~~~ Cacciari et al
(1992)~~~(d)\\
\vfill\eject

\vfill\eject
\section*{Figure captions}

\par\noindent
{\bf Figure~1.} Tracing of a portion of the co-added normalized spectrum
of X~Ari, RR~Cet and SW~And (upper, middle and lower tracing, respectively).
The spectra of X~Ari and RR~Cet were arbitrarily offset vertically to avoid
overposition of the plots.

\par\noindent
{\bf Figure~2.} Comparison between measured $EW$s of RR~Lyr and RR~Cet. The
best-fit regression line with zero constant value is overposed (see text).

\par\noindent
{\bf Figure~3.} Comparison between theoretical K92 and empirical color-\teff\
calibrations (dashed and solid lines, respectively). Filled circles are the
Population~I main sequence stars listed in Table~7.

\par\noindent
{\bf Figure~4.} Comparisons between observed H$_{\alpha}$\ profiles for X~Ari,
RR~Cet, and SW~And (thick lines), and synthetic spectra computed using K92
model atmospheres with \teff\ of 5500, 5750, 6000, 6250, and 6500~K (thin
lines: H$_{\alpha}$\ becomes stronger with increasing \teff). The spectra of
RR~Cet and SW~And were arbitrarily offset vertically to avoid overposition of
the plots. Synthesized profiles were computed assuming a surface gravity of
$\log g=2.75$\ and a metal abundance of [A/H]=-1. \teff's derived from these
fits were corrected downward by 247~K, to account for the systematic
correction determined from an analogous fit of the solar H$_{\alpha}$\
profile.

\par\noindent
{\bf Figure~5.} Comparison between the temperature stratification in the
atmosphere of a typical \RR star at minimum light as derived from K79 and K92
models, respectively.

\par\noindent
{\bf Figure~6.} (a) Plot of the residuals of abundances deduced from individual
Fe lines minus the average Fe abundances for each star, against $EW$s, for all
program stars; (b) Plot of the average residuals of panel 6a, calculated in
bins of 0.2~dex, against $EW$s (open squares). Lines represent predictions
obtained analyzing with a constant microturbulent velocity synthetic spectra
computed assuming parabolic runs of microturbulent velocity with optical depth,
with minima at $\log \tau=0$\ (solid line), $\log \tau=-1$\ (dotted line),
$\log \tau=-2$\ (short dashed line), and $\log \tau=-3$\ (long dashed line).

\par\noindent
{\bf Figure~7.} Departure coefficients for Fe~I levels from our statistical
equilibrium computations for a typical \RR model atmosphere (\teff=6200~K,
$\log g=2.75$, [A/H]=$-$1.3). Computations were made using MULTI code by
Scharmer \& Carlsson (1985), and the 61-level Fe-I model atom by Gratton et al
(1995). In panel (a) collisions with H I atoms are included ($k=31.6$), while
they are neglected in panel (b).

\par\noindent
{\bf Figure~8.} Calibration of the cross section for collisions with H~I atoms
from differences in the abundance derived from neutral and singly ionized Fe
lines. The shaded region corresponds to the range allowed for these
differences, by observational errors and uncertainties in the adopted
atmospheric parameters.

\par\noindent
{\bf Figure~9.} Comparison of the $\alpha-$elements overabundance in \RR
variables and non-variable less evolved stars in the solar neighborhood
(Edvardsson et al 1993, Nissen et al 1994; Magain 1989; Zhao \& Magain 1990;
Gratton \& Sneden 1991; Carretta et al 1995).

\par\noindent
{\bf Figure~10(a-d).} Comparison between synthetic and observed spectra for a
few stars.

\par\noindent
{\bf Figure~11.} Calibration of the cross section for collisions with H~I atoms
from: a) the multiplets at 7771-7774 and 6155-58~\AA\ of O~I, and b) the
doublets at 5893 and 5685~\AA\ of Na~I. Shaded regions correspond to the range
allowed by observational errors and uncertainties in the adopted atmospheric
parameters.

\par\noindent
{\bf Figure~12.} Corrections to the LTE abundances of a) the O~I IR triplet and
b) the Na~I lines, as a function of line strength, for model atmospheres
representative of variables having different metallicities (solar, intermediate
and low). Different curves in panel (b) correspond to the 5893~\AA\ doublet of
Na~I for \feh =0.0 (heavy solid line), \feh =$-$1.5 (heavy dotted line) and
\feh =$-$2.5 (heavy short-dashed line), respectively; and the 5688~\AA\ doublet
of Na~I for \feh =0.0 (light solid line), \feh =$-$1.5 (light dotted line) and
\feh =$-$2.5 (light short-dashed line), respectively.

\par\noindent
{\bf Figure~13.} Comparison of the [Na/Mg] abundances in \RR variables and
non-variable less evolved stars in the solar neighborhood (Edvardsson et al
1993; Nissen et al 1994; Carretta et al 1995).

\par\noindent
{\bf Figure~14.} (a) The \DS {\it vs} \feh calibration obtained from the stars
analyzed in the present paper alone (solid line); (b) The \DS {\it vs} \feh
calibration obtained by combining our \feh values (filled squares), and those
by Butler and coworkers (open squares), corrected for the systematic difference
with respect to our \feh (see Section~5.1); (c) The \DS {\it vs} \feh
calibration obtained from the total sample of field (filled and open squares)
and globular cluster RR~Lyraes (asterisks), (see Section~5.1.1). This last one
is our final recommended calibration.

\par\noindent
{\bf Figure~15.} The new $W'(K)$ {\it vs} \feh calibration (solid line),
derived from stars in common with Clementini et al (1991). The dotted line
represents the $W'(K)$ {\it vs} \feh calibration obtained in that paper.

\par\noindent
{\bf Figure 16.} M$_{V}$(RR) {\it vs} \feh using the compilation in
Table~21. Panel (a) corresponds to original M$_V$ values (Column~5 of
Table~21); panel (b) to M$_V$ values corrected to {\it p}=1.38
(Column~6 of Table~21). Different symbols refer to: field RR~Lyraes
(open and filled squares, the last ones being the objects studied in the
present paper, for them, we have used our \feh estimates); RR~Lyraes in M5 and
M92 (open triangles and open circles, respectively) according to Storm et al
(1994); and RR~Lyraes in M4 (filled triangles) according to Liu \& Janes
(1991). Field variables that are claimed to be evolved are marked as asterisks.
Best-fit regression lines obtained eliminating the evolved objects : SS Leo
and M92 stars, are plotted on the data.

\vfill\eject
\section*{Table captions}

\par\noindent
{\bf Table~1.} Program stars and observation log

\par\noindent
{\bf Table~2.} Characteristics of the co-added spectra

\par\noindent
{\bf Table~3a.} List of lines and adopted gf values and their equivalent widths
for : V445 Oph, SW And, V440 Sgr, RR Cet, UU Cet

\par\noindent
{\bf Table~3b.} List of lines and adopted gf values and their equivalent widths
for : RR Lyr, VX Her, ST Boo, VY Ser, X Ari

\par\noindent
{\bf Table~4.} Reddening of program stars, from data in the literature

\par\noindent
{\bf Table~5.} Reddening estimates from interstellar absorption features

\par\noindent
{\bf Table~6.} Metallicity values from data in the literature

\par\noindent
{\bf Table~7.} Population I main sequence stars with accurate \teff

\par\noindent
{\bf Table~8.} Effective temperatures

\par\noindent
{\bf Table~9.} \teff from H$_{\alpha}$\ profiles

\par\noindent
{\bf Table~10.} Adopted atmospheric parameters

\par\noindent
{\bf Table~11.} Dependence of abundances on atmospheric parameters

\par\noindent
{\bf Table~12.} Fe abundances

\par\noindent
{\bf Table~13.} Fe abundances from spectra at maximum light for SW And

\par\noindent
{\bf Table~14.} Abundances of $\alpha-$elements

\par\noindent
{\bf Table~15.} Overabundances of $\alpha-$elements in metal-poor \RR stars
([Fe/H]$<-1$)

\par\noindent
{\bf Table~16.} Abundances of Fe group-elements

\par\noindent
{\bf Table~17.} Abundances of heavy elements

\par\noindent
{\bf Table~18.} Overabundances of heavy elements in metal-poor \RR stars
([Fe/H]$<-1$)

\par\noindent
{\bf Table~19.} Average abundances for ionized species in metal-rich RR~Lyrae

\par\noindent
{\bf Table~20.} Results for C, O, Na and Al abundances

\par\noindent
{\bf Table~21.} M$_{V}$(RR) from B-W

\end{document}